\documentclass[aps,floatfix,twocolumn,showpacs,pra]{revtex4}
\usepackage{graphicx,bm,color}
\usepackage{amsmath,amsfonts,amssymb}
\usepackage{color}

\newcommand{\be}{\begin{equation}}
\newcommand{\ee}{\end{equation}}
\newcommand{\bea}{\begin{eqnarray}}
\newcommand{\eea}{\end{eqnarray}}
\newcommand{\rr}{\mathbf{r}}
\newcommand{\kk}{\mathbf{k}}
\newcommand{\qq}{\mathbf{q}}
\newcommand{\pp}{\mathbf{p}}

\newcommand{\bra}[1]{\langle #1|}
\newcommand{\ket}[1]{|#1\rangle}
\newcommand{\meanv}[1]{\langle #1 \rangle}
\newcommand{\meanvlr}[1]{\left\langle #1 \right\rangle}
\newcommand{\Tr}[1]{\text{Tr} \bb{#1}}
\newcommand{\bb}[1]{\left( #1 \right)}

\begin{document}
\title{Phase operators and blurring time of a pair-condensed Fermi gas}
\author{H. Kurkjian, Y. Castin, A. Sinatra}
\affiliation{Laboratoire Kastler Brossel,
Ecole Normale Sup\'erieure, UPMC and CNRS,
24 rue Lhomond, 75231 Paris Cedex 05, France}

\begin{abstract}
Due to {atomic} interactions and dispersion in the total atom number, the order parameter of a pair-condensed Fermi gas
experiences a collapse in a time that we derive microscopically. As in the bosonic case, this blurring time
depends on the derivative of the gas chemical potential with respect to the atom number and {on} the variance
of that atom number. The result is obtained first using linearized {time-dependent} Bogoliubov-de Gennes equations, then in the Random Phase Approximation, and then it is generalized to beyond mean field. 
In this framework, we construct and compare two phase operators for the paired fermionic field:
The first one, issued from our study of the dynamics, is the infinitesimal generator of adiabatic
translations in the total number of pairs. The second one is the phase operator of the amplitude of the field of pairs
on the condensate mode. We explain that these two operators differ due to the dependence of the condensate
wave function on the atom number.
\end{abstract}

\maketitle


\section{Introduction}


Long range coherence in time and space is a key property of macroscopic quantum systems
such as lasers, Bose-Einstein condensates, superfluids and superconductors.
In the case of bosonic systems, coherence derives from the macroscopic occupation of a single particle mode and can be directly visualized in an interference experiment by mixing the quantum fields extracted at two different spatial points of the system or at two different times. The interference pattern then depends on the relative phase of the two fields.

Experimental investigation of temporal coherence in Bose-Einstein condensates began right after their achievement in the laboratory \cite{Cornell1998,Kasevich2001,Bloch2002} and 
the use of their coherence properties in atomic clocks or interferometers \cite{Dunningham2005,Shin2004,Ketterle2007}, or even for the creation of entangled states, is currently a cutting-edge subject of investigation. In this respect a crucial role is played by the atomic interactions. 
On the one hand, the interactions
 limit the coherence time causing an initially well defined phase or relative phase to blur {in a finite size system}
 {\cite{Sols1994,Walls1996,Lewenstein1996,CastinDalibard1997,Villain1997,Berrada2013,Will2010}}; on the other hand, at shorter times, 
coherent phase dynamics in presence of interactions allows for generation of spin squeezed states \cite{Oberthaler2010,Treutlein2010} that opens the way to quantum metrology \cite{Ueda1993,Wineland1994,
Sorensen:2001,Frontiers:2012,Sinatra2012}. 

Let us now turn to the case of fermions.
Cold fermionic  gases have been widely studied in the last decade \cite{Varenna_livre,Stringari2008,Zwerger2012}. With respect to bosons, fermions have the advantage that the interaction strength can be changed by the use of Feshbach resonances without introducing significant losses in the system. Across these resonances the $s$-wave scattering length characterizing the short range interactions in the cold gas can be ideally  changed from $-\infty$ to $+\infty$.  Using the same physical system, different interaction regimes can be accessed, ranging from a BCS (Bardeen-Cooper-Schrieffer) superfluid of weakly bound Cooper pairs,
when $a$ is small and negative, to a condensate of tightly bound dimers behaving like bosons, when $a$ is small and positive. In between, the strongly interacting unitary gas is obtained when
$a$ diverges \cite{Varenna,Randeria2014}.
{Recent progress in the experiments} made it possible to observe the coherence and superfluidity of these Fermi gases  \cite{Ketterle2005,Grimm:2013}, to study with high precision {their} thermodynamics in the different interaction regimes \cite{Salomon2010,Salomon_Nature2010,Felix,Zwierlein} and to perform an interference experiment between two independent condensates of dimers \cite{Kohstall2011}. 

In the near future we expect the experimental studies to extend to coherence properties {e.g. along the lines
of \cite{Altman2004,CarusottoCastin2005}} and this motivates a theoretical study of phase dynamics in paired fermionic systems. The scope of this paper is to provide an analysis of this problem, at zero temperature.

Let us consider an unpolarized Fermi gas with two internal states
$\uparrow$ and $\downarrow$, in presence of weak attractive interactions between fermions in different 
internal states. At zero temperature, a macroscopic number of $\uparrow\downarrow$ pairs of fermions condense in the same two-body wave function. Long range order and coherence properties then show up in the \textit{two-body correlations} that are in principle measurable \cite{Altman2004,CarusottoCastin2005}. 

To investigate temporal coherence of the pair-condensed Fermi system, in a first stage, we determine the time evolution of the \textit{order parameter} of a state that is a {coherent} superposition of different particle numbers. At the mean field level, the broken symmetry state is simply the ground state of the gas in the BCS theory \cite{BCS1957}. 
Interactions and an initial dispersion in the particle number cause a blurring of the phase and the collapse of the order parameter of the BCS gas, after a time that we derive analytically. This is done in section \ref{sec:collapse}. Loss of coherence is described in a low-energy subspace of the linearized equations of motion by a zero-energy mode \cite{Anderson1958} and an anomalous mode \cite{Ripka1985} excited respectively by phase and particle number translations. We derive these modes and the related phase and number operators explicitly and express the phase blurring time in terms of the particle number fluctuations and of thermodynamic quantities of the gas. We show that the same microscopic expression of the blurring time is obtained using the Random Phase Approximation put forward in \cite{Anderson1958}. A similar symmetry breaking approach, based on the linearized treatment of quantum fluctuations, was introduced for bosonic phase dynamics in \cite{Lewenstein1996}. {We stress that the collapse we are interested in is a finite size effect and the blurring time
we find diverges in the thermodynamic limit.}

In section \ref{sec:phases} we make some general considerations about the definition of a phase operator in the fermionic case and we examine two possible candidates for this operator. We find that the phase operator derived in section \ref{sec:collapse} in the dynamical study of the order parameter is a generator of \textit{adiabatic} translations of the number of particles of the gas, that is it increases the particle number while leaving the system in its many-body ground state. A second natural definition of the phase operator, that we call the phase operator of the condensate of pairs, is associated to the amplitude of the field of pairs $\hat{\psi}_\downarrow(\rr) \hat{\psi}_\uparrow(\rr')$ on the condensate wave function, defined as the macroscopically populated mode of the two-body density matrix.
We show that the two phase operators {\it differ} if the condensate wave function depends on the total number of particles. 
Although not pointed out at that time, this difference was already present in the studies of bosonic phase dynamics
\cite{Lewenstein1996,Villain1997,CastinDum1998,EPL}.

In Sec. \ref{sec:beyond} we extend our results for the blurring time and the phase operator beyond the BCS theory. This is done in two ways. First, we go beyond $U(1)$ symmetry breaking: 
although they are not appropriate to describe the state of an isolated gas, as they cannot be prepared experimentally, broken symmetry states can be given a precise physical meaning when dealing with a bi-partite system with a well defined relative phase \cite{CastinDalibard1997,CarusottoCastin2005}.
In Sec. \ref{subsec:restoringsymmetry} we restore the symmetry by considering a mixture of broken symmetry states,
and we relate the order parameter to correlation functions. 
Second, in Sec. \ref{subsec:beyondmeanfield}
we go beyond the mean field regime by replacing the BCS ansatz by a {coherent} superposition of the exact ground states for different number of particles. {We have chosen to postpone this section giving a general result after the microscopic derivations 
based on the mean field BCS theory that are useful to set the stage and become familiar with the problem. Nevertheless subsection \ref{subsec:beyondmeanfield} is self-contained, and the reader willing to avoid
all technicalities might go to this subsection directly.}
 
Finally, in Sec. \ref{sec:hintsforanexperiment} we sketch an experiment in which such state would be prepared and which would allow  the observation of a Gaussian decay of the time-correlation function. We conclude in section \ref{sec:conclusion}.


\section{Collapse of the BCS order parameter}


\label{sec:collapse}

{In this section we show that the phase of the order parameter of a gas initially prepared in the $U(1)$ 
symmetry-breaking BCS state spreads in time, causing the order parameter to collapse. To this aim, we use linearized equations of motion both at the {``classical"} mean field level and at the quantum level (RPA).  The two approaches yield equivalent equations of motions for small fluctuations, either classical or quantum, of the dynamical variables. However while the quantum fluctuations of the initial state are build-in in the quantum theory, we need an \textit{ad hoc} probability distribution of the classical fluctuations to reproduce the quantum behavior with the ``classical" theory.}


\subsection{Hamiltonian}


We consider a gas of fermions in two internal states $\uparrow$ and $\downarrow$, in the grand canonical ensemble of chemical potential $\mu$, in a cubic lattice model of step $b$ with periodic boundary conditions in $[0,L]^3$.
The fermions have on-site interactions characterized by the bare coupling constant $g_0$. The grand canonical Hamiltonian of this system is given by
\begin{multline}
\label{eq:Hamiltonian}
\hat{H}= b^3  \sum_{\rr, \sigma} \hat{\psi}_\sigma^\dagger(\rr) \left( - \frac{\hbar^2}{2m}  \Delta_{\rr} -\mu\right)  \hat{\psi}_\sigma(\rr) \\
+ g_0 b^3 \sum_{\rr} \hat{\psi}_\uparrow^\dagger(\rr) \hat{\psi}_\downarrow^\dagger(\rr)\hat{\psi}_\downarrow(\rr){\hat{\psi}_\uparrow}(\rr)
\end{multline}
where the single particle discrete laplacian on the lattice has plane waves $e^{i\kk \cdot \rr}$ as eigenvectors with eigenvalues $\hbar^2 k^2/2m$ and the field operators obey discrete anti-commutation relations such as
$\{ \hat{\psi}_\sigma(\rr) , \hat{\psi}_{\sigma'}^\dagger({\rr'}) \} = \delta_{\sigma \sigma'} \delta_{\rr \rr'}/b^3$, and $\sigma, \sigma'=\uparrow {\rm or} \downarrow$.
The bare coupling constant $g_0$ is adjusted to reproduce the scattering length $a$ of the true interaction potential \cite{CastinMora2003,LesHouches2004,Burovski2006,Varenna,Pricoupenko2007,Juillet2007}:
\begin{equation}
\label{eq:g0}
\frac{1}{g_0} = \frac{m}{4 \pi \hbar^2 a} - \int_{\rm FBZ} \frac{d^3k}{(2\pi)^3} \frac{m}{\hbar^2 k^2} 
\end{equation} 
where FBZ is the first Brillouin zone $[-\pi/b,\pi/b[^3$ of the lattice.


\subsection{Reminder of the BCS theory}


The BCS theory is based on the introduction of the ansatz
\begin{equation}
| \psi_{\rm BCS} \rangle =\mathcal{N} e^{\gamma \hat{C}^\dagger} | 0 \rangle
\label{eq:BCSstate}
 \end{equation}
where $\gamma$ is a complex number and $\hat{C}^\dagger$ creates a pair of $\uparrow\downarrow$ fermions in a wave function $\varphi$. In Fourier space
\begin{equation}
\hat{C}^\dagger= \sum_{{\kk}} \varphi_\kk \hat{a}_{{\kk} \uparrow}^\dagger \hat{a}_{-{\kk} \downarrow}^\dagger
 \end{equation}
where $\hat{a}_{\kk \sigma}^\dagger$ creates a fermion of wave vector $\kk$ in spin state $\sigma$ and obeys
the usual anti-commutation relations. For the purpose of this work, it is sufficient to restrict to pairs of zero
total momentum, as this will describe both the initial ground BCS state and the relevant fluctuations for phase dynamics. 
To parametrize the BCS state we use the complex $V_\kk$ coefficients:
\begin{equation} 
V_\kk=-\frac{\gamma \varphi_\kk}{\sqrt{1+|\gamma\varphi_\kk|^2}}
\label{eq:defVk}
\end{equation}
which can be interpreted as the probability amplitudes of finding a pair with wave vectors $\kk$ and $-\kk$:
\bea
|V_\kk|^2 &=& \langle \hat{a}_{{\kk} \uparrow}^\dagger \hat{a}_{-{\kk} \downarrow}^\dagger \hat{a}_{-{\kk} \downarrow} \hat{a}_{{\kk} \uparrow} \rangle_{\rm BCS} \nonumber \\
&=& 
\langle \hat{a}_{{\kk} \uparrow}^\dagger \hat{a}_{{\kk} \uparrow}  \rangle_{\rm BCS}
=\langle \hat{a}_{{-\kk} \downarrow}^\dagger \hat{a}_{{-\kk} \downarrow}  \rangle_{\rm BCS}
\eea
where the notation $\meanv{\ldots}_{\rm BCS}$ means that the average value is taken in the BCS state \eqref{eq:BCSstate}.
We rewrite this state using the parameters \eqref{eq:defVk} in the usual form (up to a different sign convention)
\begin{equation}
\label{eq:usual_BCS}
| \psi_{\rm BCS} \rangle =\prod_\kk \bb{U_\kk-V_\kk \hat{a}_{{\kk} \uparrow}^\dagger \hat{a}_{-{\kk} \downarrow}^\dagger} \ket{0}
 \end{equation}
where $U_\kk$ defined by
\be
U_\kk\equiv\sqrt{1-|V_\kk|^2}
\label{eq:Uk}
\ee
is real and positive.
The BCS ansatz breaks the $U(1)$ symmetry and has a non-zero order parameter $\Delta$ 
\be
\label{eq:defDelta}
\Delta \equiv g_0 \meanv{\hat{\psi}_\downarrow \hat{\psi}_\uparrow}_{\rm BCS}  
 = - \frac{g_0}{L^3} \sum_\kk V_\kk U_\kk \neq 0 
\ee

The BCS ground state $| \psi^0_{\rm BCS} \rangle$ is obtained by minimizing the energy functional 
\begin{equation}
\label{eq:fonctionnelle}
E=\meanv{\hat{H}}_{\rm BCS}
\end{equation}
treated as a \textit{classical} Hamiltonian with respect to the complex parameters $V_\kk$
 \begin{equation}
\frac{\partial E}{\partial V_\kk}\bigg|_{V_\kk=V_\kk^0}=\frac{\partial E}{\partial V_\kk^{*}}\bigg|_{V_\kk=V_\kk^0}=0
\end{equation}
Explicitly
\be
E=g_0 L^3 \rho_\uparrow \rho_\downarrow + \frac{L^3}{g_0}|\Delta|^2+
\sum_\kk 2 \bb{\frac{\hbar^2 k^2}{2m} -\mu} |V_\kk|^2  
\label{eq:Eclass}
\ee
where the average density of spin $\sigma$ particles, 
\be
\rho_\sigma \equiv \meanv{\hat{\psi}_{\sigma}^\dagger \hat{\psi}_{\sigma}}_{\rm BCS}=\frac{1}{L^3} \sum_\kk |V_\kk|^2  \label{eq:rhoup} 
\ee
{is here spin-independent}.
This minimization leads to
{
\begin{equation}
2 \xi_\kk U_\kk^{0} V_\kk^{0} = \Delta_0 [(U_\kk^{0})^2 - (V_\kk^{0})^2 ]
\label{eq:utile3}
\end{equation}
whose solution with positive $V_\kk^{0}$ is}
 \begin{equation}
 \label{eq:gap}
V_\kk^0 = \sqrt{\frac{1}{2} \left( 1-\frac{\xi_\kk}{\epsilon_\kk} \right)}
\end{equation}
where $\xi_\kk=\frac{\hbar^2 k^2}{2m} - \mu + g_0 \rho_\uparrow^0$ is the kinetic energy shifted by the chemical potential and corrected by the mean-field energy, and 
\be
\epsilon_\kk=\sqrt{\xi_\kk^2 + \Delta_0^2}
\label{eq:epsilonk}
\ee
is the energy of the BCS pair-breaking excitations. The \textit{gap} $\Delta_0$ is the ground state value of the order parameter \eqref{eq:defDelta} 
\be
\Delta_0\equiv g_0 \meanv{\hat{\psi}_\downarrow \hat{\psi}_\uparrow}_{\rm 0}
\ee
where the notation $\meanv{\ldots}_{\rm 0}$ means that the average value is taken in the BCS ground state $| \psi^0_{\rm BCS} \rangle$.
The parameters of $| \psi^0_{\rm BCS} \rangle$ depend on the unknowns $\Delta_0$ and 
the average total density $\rho^0=\rho_\uparrow^0+\rho_\downarrow^0$, implicitly related to $\mu$ and to the scattering length $a$ by
\begin{align}
\label{eq:BCS1}
\rho^0 &=  \frac{1}{L^3} \sum_\kk \left( 1 - \frac{\xi_\kk}{\epsilon_\kk}\right) \\
\label{eq:BCS2}
- \frac{1}{g_0} &= \frac{1}{L^3} \sum_\kk  \frac{1}{2\epsilon_\kk }
\end{align}
Equations (\ref{eq:BCS1}) and (\ref{eq:BCS2}) are obtained using the mean particle number equation 
\begin{equation}
{\bar N} \equiv \langle \hat{N}\rangle_0=2 \sum_\kk (V_\kk^0)^2
\label{eq:nbar}
\end{equation}
in addition to (\ref{eq:g0}) and (\ref{eq:gap}). Equation \eqref{eq:BCS2} is called the {\it gap equation}.


\subsection{{Time-dependent Bogoliubov-de Gennes} approach: zero frequency mode, anomalous mode and phase dynamics}


\label{sec:semiclassical}


\subsubsection{Linearized time-dependent Bogoliubov-de Gennes equations}


We now consider a nonstatic BCS state $\ket{\psi_{\rm BCS}(t)}$, of parameters 
\begin{equation}
V_\kk(t)=V_\kk^0 + \delta V_\kk(t)
\end{equation}
\textit{The time-dependent BCS equations} (or Bogoliubov-de Gennes equations) arise from the minimization of the action
 \begin{equation}
 S= \int_{t_i}^{t_f} dt \left\{ \frac{i \hbar}{2} \left( \langle \psi(t) |\frac{d}{dt}|\psi(t)\rangle - {\rm c.c.} \right) - \langle \psi(t) |\hat{H}|  \psi(t) \rangle \right\}
\end{equation}
Choosing $U_\kk(t)$ as in \eqref{eq:Uk} real at all times, this gives \cite{Ripka1985}
 \begin{equation}
\label{eq:tdBCS} 
i \hbar \frac{dV_\kk}{dt}= \frac{\partial E}{\partial V_\kk^{*}} \qquad i \hbar \frac{dV_\kk^*}{dt}= - \frac{\partial E}{\partial V_\kk}
\end{equation}
These equations can be linearized around the BCS ground state for small perturbations $\delta V_\kk$.
Introducing the operator $\mathcal{L}$ such that
\begin{equation}
\label{eq:timedependent}
i \hbar \frac{d}{dt}
\begin{pmatrix} {\delta V_\kk}\\{\delta V_\kk^*} \end{pmatrix}_{\kk \in \mathcal{D}}
= \mathcal{L} \begin{pmatrix} {\delta V_\kk}\\{\delta V_\kk^*}  \end{pmatrix}_{\kk \in \mathcal{D}}
\end{equation}
where $\mathcal{D}$ contains all possible values of the single particle wave vector, {$\mathcal{D}=[-\frac{\pi}{b},\frac{\pi}{b}[^3\cap\bb{\frac{2\pi}{L}\mathbb{Z}^3}$}, we can write $\mathcal{L}$ in a block form using the derivatives of $E$
taken in the ground BCS state:
\begin{eqnarray}
&\mathcal{L}= \begin{pmatrix} A & B \\  -B^* & -A^* \end{pmatrix} \label{eq:Lblock}\\
&A_{\kk \qq}=\displaystyle \frac{\partial^2 E}{\partial V_\qq \partial V_\kk^*} \qquad B_{\kk \qq}= \frac{\partial^2 E}{\partial V_\qq^* \partial V_\kk^*} \notag
\end{eqnarray}
The matrix $A$ is hermitian and the matrix $B$ is symmetric, we give their explicit expressions in Appendix \ref{app:equations_semi}.
The operator $\mathcal{L}$ gives access to the time evolution of a given perturbation:
\begin{equation}
\begin{pmatrix} {\delta V_\kk} \\{\delta V_\kk^*} \end{pmatrix}_{\kk \in \mathcal{D}} (t_f)= \exp \left[\frac{-i \mathcal{L} (t_{f} - t_{i}) }{\hbar}\right]\begin{pmatrix} {\delta V_\kk} \\{\delta V_\kk^*} \end{pmatrix}_{\kk \in \mathcal{D}} (t_i)
\end{equation}
and to the energy difference between the perturbed state and the BCS ground state
up to second order in the perturbation:
\begin{equation} 
\label{eq:energie}
E[(V_\kk,V_\kk^*)_{\kk \in \mathcal{D}}]=E_{0} + \frac{1}{2} \begin{pmatrix} {\delta V_\kk^*} & {\delta V_\kk} \end{pmatrix}_{\kk \in \mathcal{D}} \sigma_z \mathcal{L}\begin{pmatrix} {\delta V_\kk} \\{\delta V_\kk^*} \end{pmatrix}_{\kk \in \mathcal{D}}
\end{equation}
where $E_0=\meanv{\hat{H}}_0$ and $\sigma_x = \begin{pmatrix} 0 & \openone \\  \openone & 0 \end{pmatrix}$, $\sigma_z = \begin{pmatrix} \openone & 0 \\  0 & -\openone \end{pmatrix}$ are {block} Pauli matrices.
Note that the matrix $\sigma_z \mathcal{L}$ is hermitian by construction and it is non negative since $E_0$ is the ground 
state BCS energy.
In full analogy with previous results for bosons \cite{CastinDum1998}, our choice of canonically conjugate variables leads to a highly symmetric linearized evolution operator. The \textit{symplectic} symmetry 
\begin{eqnarray} 
\label{eq:symplectique}
\sigma_z \mathcal{L} \sigma_z =\mathcal{L}^\dagger
\end{eqnarray}
ensures that the eigenvectors of $\mathcal{L}^\dagger$ are equal to those of $\mathcal{L}$ multiplied by $\sigma_z$. The \textit{time reversal} symmetry
\begin{equation}
\sigma_x \mathcal{L} \sigma_x= -\mathcal{L}^*
\end{equation}
ensures that for each eigenvector $(a,b)$ of $\mathcal{L}$ with eigenvalue $\epsilon$, $(b^*,a^*)$ is also an eigenvector of $\mathcal{L}$ with eigenvalue $-\epsilon^*$.


\subsubsection{Zero-energy subspace}


\label{sec:energie_nulle}

We concentrate here on the zero-energy subspace of $\mathcal{L}$ where zero temperature phase dynamics occurs.


\paragraph{Zero energy mode}


Due to the $U(1)$ symmetry of the Hamiltonian, the mean energy of a BCS state does not depend on the phase of the parameter $\gamma$ in \eqref{eq:BCSstate}, that is it is invariant by the transformation
\begin{eqnarray}
\gamma  & \to & \gamma e^{iQ} \notag \\
V_\kk  & \to & V_\kk e^{iQ} \notag \\
V_\kk^* & \to & V_\kk^* e^{-iQ} \notag \\
\end{eqnarray}
Consequently the classical Hamiltonian $E$ \eqref{eq:Eclass} is not affected by a global phase change of the BCS ground state parameters {$V_\kk^0$, $E[(V_\kk^0e^{iQ},V_\kk^0e^{-iQ})_{\kk \in \mathcal{D}}]=E_0$}. 
From \eqref{eq:energie} this implies that the perturbation linearized for $Q \ll 1$,
\begin{equation}
\vec{e}_n \equiv \begin{pmatrix} {i V_\kk^0} \\{-i V_\kk^{0}} \end{pmatrix}_{\kk \in \mathcal{D}}
\label{eq:vec_e}
\end{equation}
 is a  zero-energy (null energy) mode
\begin{equation}
 \mathcal{L} \vec{e}_n = \vec{0}
 \label{eq:null_mode}
\end{equation}
Alternatively we can consider the continuous family $Q \mapsto (V_\kk^0e^{iQ},V_\kk^0e^{-iQ})_{\kk \in {\cal D}}$. 
Each element
of this family is a time independent solution of \eqref{eq:tdBCS}.  For $Q$ infinitesimal, the difference with respect to the
$(V_\kk^0,V_\kk^0)_{\kk \in {\cal D}}$ member of the family is then a zero frequency solution of \eqref{eq:timedependent}.

The vector \eqref{eq:vec_e} is equal to its time-reversal symmetric $\vec{e}_n= \sigma_x \vec{e}_n^{\,*}$, and does not span the full zero-energy subspace. We will obtain the missing vector in the following paragraph.


\paragraph{Anomalous mode}


After phase translations, one naturally turns to mean particle number translations. By \textit{adiabatically} varying the chemical potential $\mu$ of the gas (i.e. by changing the mean number of particles continuously following the BCS ground state), we will prove the existence of an \textit{anomalous mode} with the properties
\begin{eqnarray}
\mathcal{L} \vec{e}_a &=& - 2i \frac{d\mu}{d\bar{N}}  \vec{e}_n \label{eq:anomalousmode} \\
\mathcal{L}^2 \vec{e}_a &=&\vec{0} \label{eq:nihil}
\end{eqnarray}

Let us introduce $V_\kk^0(\tilde{\mu})\in \mathbb{R}^+$, the parameters of the ground state BCS solution corresponding to a chemical potential $\tilde{\mu}$. Within the BCS ansatz, they minimize the mean value of 
$\hat{H}+(\mu-\tilde{\mu})\hat{N}$ with $\hat{H}$ given by \eqref{eq:Hamiltonian}.
We then consider the family of time-dependent BCS parameters 
\be
\tilde{\mu} \mapsto V_\kk(\tilde{\mu},t) \equiv  V_\kk^0(\tilde{\mu}) e^{-2i (\tilde{\mu}-\mu)t/\hbar}
\label{eq:family_tilde}
\ee
We will show later on that each element of this family is a solution of the time dependent BCS equations
\eqref{eq:tdBCS} for a chemical potential $\mu$, the phase factor in \eqref{eq:family_tilde} precisely compensating the mismatch between the two chemical potentials  $\mu$ and $\tilde{\mu}$. 
By linearizing \eqref{eq:family_tilde} for a small value of $\tilde{\mu}-\mu$, we obtain the deviations
\be
 \begin{pmatrix} {\delta V_\kk} \\{\delta V_\kk^*} \end{pmatrix}_{\kk \in \mathcal{D}} \!\!\!\!\!\! = 
(\tilde{\mu}-\mu)\left[-\frac{2t}{\hbar}\begin{pmatrix} {iV_\kk^0(\mu)} \\{-iV_\kk^0(\mu)} \end{pmatrix} +
 \begin{pmatrix} {\frac{d}{d\mu} V_\kk^0({\mu})} \\{\frac{d}{d\mu} V_\kk^0({\mu})} \end{pmatrix} \right]_{\kk \in \mathcal{D}}
\label{eq:deviation}
\ee
which must be a solution of the linearized time dependent BCS equations \eqref{eq:timedependent}.
Using the expression \eqref{eq:vec_e} for the zero-energy mode, this explicitly gives the announced anomalous mode \eqref{eq:anomalousmode}
\begin{equation}
\vec{e}_a = \frac{d\mu}{d\bar{N}} \begin{pmatrix} { \frac{d}{d\mu} V_\kk^0} \\{ \frac{d}{d\mu} V_\kk^{0}} \end{pmatrix}_{\kk \in \mathcal{D}} 
\end{equation}

Let us now show as promised that the family \eqref{eq:family_tilde} is a solution of the time dependent BCS equations
\eqref{eq:tdBCS} for a chemical potential $\mu$. A first way is to remark that if $|\tilde{\psi}\rangle$ is a solution
of the time dependent Schr\"odinger equation of Hamiltonian $\hat{H}+(\mu-\tilde{\mu})\hat{N}$, then
$e^{-i(\tilde{\mu}-\mu)t \hat{N}/\hbar}|\tilde{\psi}\rangle$ is a solution
of the time dependent Schr\"odinger equation of Hamiltonian $\hat{H}$. By the application of this unitary 
transformation to the ground-state BCS solution in the usual form \eqref{eq:usual_BCS} for a chemical potential $\tilde{\mu}$ and using 
\be
e^{-i(\tilde{\mu}-\mu)t \hat{N}/\hbar} \hat{a}_{\kk \uparrow}^\dagger \hat{a}_{-\kk \downarrow}^\dagger  
e^{i(\tilde{\mu}-\mu)t \hat{N}/\hbar} = e^{-2i(\tilde{\mu}-\mu)t/\hbar}\hat{a}_{\kk \uparrow}^\dagger \hat{a}_{-\kk \downarrow}^\dagger  
\ee
we get the announced result.
Alternatively one can directly inject the form \eqref{eq:family_tilde} into the time dependent BCS equations
\eqref{eq:tdBCS}, which also gives the announced result. Introducing $\tilde{E}$ as the classical Hamiltonian for a chemical potential $\tilde{\mu}$
one indeed has from \eqref{eq:Eclass}
\begin{multline}
E[(V_\kk(\tilde{\mu},t),V_\kk^*(\tilde{\mu},t))_{\kk \in {\cal D}}] = \tilde{E}[(V_\kk(\tilde{\mu},t),V_\kk^*(\tilde{\mu},t))_{\kk \in {\cal D}}]\\
+ 2 (\tilde{\mu}-\mu) \sum_\kk |V_\kk(\tilde{\mu},t)|^2
\end{multline}
Furthermore the derivatives $\partial \tilde{E}/\partial V_\kk^*$ and $\partial \tilde{E}/\partial V_\kk$
vanish when evaluated in the point $(V_\kk(\tilde{\mu},t),V_\kk^*(\tilde{\mu},t))_{\kk \in {\cal D}}$ since
this point coincides with the ground state point $(V_\kk^0(\tilde{\mu}),V_\kk^0(\tilde{\mu}))_{\kk \in {\cal D}}$
up to a global phase factor.


\paragraph{Dual vectors of $\vec{e}_n$ and $\vec{e}_a$}


An arbitrary fluctuation of components $\delta V_\kk$ and $\delta V_\kk^*$ can be expanded on the basis formed 
by the anomalous mode $\vec{e}_a$ and the eigenvectors of ${\cal L}$ including the zero-energy mode $\vec{e}_n$
and the excited eigenmodes $\vec{e}_\lambda$.
To obtain the coefficients of such an expansion, we introduce the dual basis (also called adjoint basis) formed by
$\vec{d}_a$, $\vec{d}_n$ and the duals of the excited modes $\vec{d}_\lambda$ such that
\be
\vec{d}_i^{\,*} \cdot \vec{e}_j =\delta_{i j}  \quad \mbox{where} \quad i,j=n,a,\lambda
\label{eq:duals}
\ee
We now calculate explicitly $\vec{d}_n$ and $\vec{d}_a$ using the symplectic symmetry (\ref{eq:symplectique}).
By taking the hermitian conjugate of \eqref{eq:null_mode} and of \eqref{eq:anomalousmode} and using (\ref{eq:symplectique}) we obtain $\forall \vec{x}$:
\begin{eqnarray}
&& ( \sigma_z \vec{e}_n )^* \cdot \mathcal{L} \vec{x} =0 \label{eq:i} \\ 
&& ( \sigma_z \vec{e}_a )^* \cdot \mathcal{L}\vec{x} = 2i \frac{d\mu}{d\bar{N}} (\sigma_z \vec{e}_n)^* \cdot  \vec{x}
\label{eq:ii}
\end{eqnarray}
This suggests that the dual vectors of $\vec{e}_n$ and $\vec{e}_a$ are obtained by the action of $\sigma_z$
on $\vec{e}_a$ and $\vec{e}_n$ respectively. Indeed we obtain 
\bea
\vec{d}_n &=& 2i \sigma_z  \vec{e}_a = 2 \frac{d\mu}{d{\bar N}} 
 \begin{pmatrix} {i \frac{d}{d\mu} V_\kk^0} \\{- i\frac{d}{d\mu} V_\kk^0} \end{pmatrix}_{\kk \in {\cal D}}
 \label{eq:dn} \\
\vec{d}_a &=& - 2i \sigma_z  \vec{e}_n = 2 
 \begin{pmatrix} { V_\kk^0} \\{ V_\kk^0} \end{pmatrix}_{\kk \in {\cal D}} 
 \label{eq:da} 
\eea
To check that ${\vec{d}_a^{\,*}} \cdot \vec{e}_\lambda=0$ we simply take $\vec{x}=\vec{e}_\lambda$ in \eqref{eq:i}.
Further taking $\vec{x}=\vec{e}_\lambda$ in \eqref{eq:ii} gives ${\vec{d}_n^{\,*}} \cdot \vec{e}_\lambda=0$.
Taking $\vec{x}=\vec{e}_a$ in \eqref{eq:i} and using \eqref{eq:anomalousmode} gives $\vec{d}_a^{\,*} \cdot \vec{e}_n=0$.
The last orthogonality relation $\vec{d}_n^{\,*} \cdot \vec{e}_a=0$ can be checked by direct substitution. 
Finally the normalization conditions  $\vec{d}_a^{\,*} \cdot \vec{e}_a=\vec{d}_n^{\,*} \cdot \vec{e}_n=1$ result from
the relation
\be
\sum_\kk V_\kk^0 \frac{d}{d\mu} V_\kk^0 = \frac{1}{4} \frac{d {\bar N}}{d\mu}
\label{eq:dNdmu}
\ee
obtained from \eqref{eq:nbar} by a derivation with respect to $\mu$.


\subsubsection{Phase variable and phase dynamics}


We expand a classical fluctuation over the modes introduced in the previous subsubsection:
\begin{equation}
\label{eq:expansion}
\begin{pmatrix} {\delta V_\kk} \\{\delta V_\kk^{*}} \end{pmatrix}_{\kk \in {\cal D}} = 
P \vec{e}_a +  Q \vec{e}_n + \sum_\lambda C_\lambda \vec{e}_\lambda
\end{equation}
The time dependent coefficients $P$ and $Q$ of the anomalous and zero-energy modes are determined by projection upon their dual vectors:
\begin{eqnarray}
P &=& 2  \sum_\kk V_\kk^{0} (\delta V_\kk +  \delta V_\kk^*) = \delta N \label{eq:P}\\
Q &=& -2i \frac{d\mu}{d\bar{N}} \sum_\kk \bb{\frac{d}{d\mu} V_\kk^{0}} (\delta V_\kk - \delta V_\kk^*) 
\label{eq:Q}
\end{eqnarray}
{and are real quantities.}
We interpret $P$ as the \textit{classical particle number fluctuation}, from linearization of {\eqref{eq:rhoup}}. 
To interpret Q we consider the infinitesimal phase translation: 
\begin{eqnarray}
\gamma  \to \gamma e^{i \delta \phi} \qquad \delta V_\kk = i \delta \phi V_\kk^0
\end{eqnarray}
Inserting such fluctuation in \eqref{eq:Q} and using \eqref{eq:dNdmu}  gives $Q= \delta\phi$.
For reasons that will  become clear in the next section, we call $Q$ the \textit{classical adiabatic phase}.

The two quantities $P$ and $Q$ are canonically conjugate classical variables. Defining the Poisson brackets
as
\be
\left\{ X,Y \right\} =  \frac{1}{i\hbar} \sum_\kk \frac{\partial X}{\partial (\delta V_\kk)}\frac{\partial Y}{\partial (\delta V_\kk^*)}-\frac{\partial X}{\partial (\delta V_\kk^*)}\frac{\partial Y}{\partial (\delta V_\kk)} 
\ee
so that $\frac{d}{dt} X = \{X,E\}$, we obtain
\be
\left\{ Q,P \right\}= -\frac{2}{\hbar}
\ee
Inserting the modal decomposition (\ref{eq:expansion}) in the quadratized Hamiltonian \eqref{eq:energie},
we find that the $P$ and $Q$ variables appear only {\it via} a term proportional to $P^2$
\be
E[(V_\kk,V_\kk^*)_{\kk \in {\cal D}}] = E_0 + \frac{1}{2}\frac{d\mu}{d\bar{N}}   P^2 + \ldots
\ee
where the ``$\ldots$" only involve the excited modes amplitudes $C_\lambda$. 
This implies that $P$ is a constant of motion and that $Q$ has a ballistic trajectory\footnote{$P$ and $Q$ have vanishing Poisson brackets with the excited modes amplitudes $C_\lambda$ as
$\{P,C_\lambda \}=(2/\hbar)\,\vec{d}_\lambda^* \cdot \vec{e}_n=0$ and
$\{Q,C_\lambda \}=(-2/\hbar)\,\vec{d}_\lambda^* \cdot \vec{e}_a=0$.}:
\bea
\frac{d}{dt}{P}  &=& \{P,E\}=  0 \label{eq:Pdot} \\
\frac{d}{dt}{Q}  &=& \{Q,E\}=   -\frac{2}{\hbar} \frac{d\mu}{d\bar{N}} P 
\label{eq:Qdot}
\eea
If $P=\delta N$ fluctuates from one realization to the other the slope of the classical phase evolution changes from shot to shot, and the overall phase distribution spreads out ballistically. For a classical distribution having zero first moments
for $P$ and $Q$ one has:
\begin{equation}
\begin{aligned}  \meanv{Q}_{\rm cl}(t) &= -\frac{2}{\hbar} \frac{d\mu}{d\bar{N}} \meanv{P}_{\rm cl} t = 0 \notag \\
 \langle \left[Q(t)-Q(0)\right]^2 \rangle_{\rm cl} &=  \frac{4t^2}{\hbar^2}\left(\frac{\text{d}\mu}{\text{d}\bar{N}}\right)^2
 \langle P^2 \rangle_{\rm cl}\    \end{aligned} 
\end{equation}
If we choose our classical probability distribution to mimic quantum fluctuations in the ground state of the BCS theory,
thus with $\langle P^2 \rangle_{\rm cl}=\meanvlr{(\hat{N}-\bar{N})^2}_0$, we obtain from our
classical approach a phase blurring time scale
\begin{equation}
t_{\rm br} = \hbar \left[ \textrm{Var \,} \hat{N} \left(\frac{d \mu}{d\bar{N}} \right)^2 \right]^{-1/2}
\label{eq:tbr}
\end{equation}


\subsection{Quantum approach: adiabatic phase operator}


\label{sec:RPA}

In this {subsection}, we use a fully quantum approach to quantize the conjugate phase and number variables of the classical approach of the previous subsection. The quantum approach  uses Anderson's Random-Phase Approximation (RPA) \cite{Anderson1958} treatment of the interaction term of the full Hamiltonian (\ref{eq:Hamiltonian}) to derive linearized equations of motion directly for the two-body operators, rather than for classical perturbations. 

We introduce the quadratic operators
\begin{align}
 \hat{n}_{\kk\uparrow}^\qq &= \hat{a}_{\kk+\qq \uparrow}^\dagger \hat{a}_{\kk \uparrow},   &\hat{\bar{n}}_{\kk\downarrow}^{\qq} &= \hat{a}_{-\kk \downarrow}^\dagger \hat{a}_{-\kk-\qq \downarrow} \notag \\
 \hat{d}_\kk^\qq &= \hat{a}_{-\kk-\qq \downarrow} \hat{a}_{\kk \uparrow},   &\hat{\bar{d}}_\kk^{\qq} &= \hat{a}_{\kk+\qq \uparrow}^\dagger \hat{a}_{-\kk \downarrow}^\dagger
\label{eq:quadratic_operators}
\end{align}
The equations of motion of these operators {in the Heisenberg picture} involve quartic terms, for example for 
$\hat{n}_{\kk \uparrow}^\qq$:
\begin{multline}
-i \hbar \frac{d}{dt}\hat{n}_{\kk  \uparrow}^\qq = [\hat{H},\hat{n}_{\kk  \uparrow}^\qq] = (\xi_{\kk+\qq}-\xi_\kk)\hat{n}_{\kk  \uparrow}^\qq  \\ + \frac{g_0}{L^3} \sum_{\kk',\pp} \left( \hat{a}_{\kk' \uparrow}^\dagger \hat{a}_{-\kk'+\pp \downarrow}^\dagger \hat{a}_{-\kk-\qq+\pp \downarrow} \hat{a}_{\kk \uparrow} \right.  \\  -   \left.  \hat{a}_{\kk +\qq \uparrow}^\dagger \hat{a}_{-\kk+\pp \downarrow}^\dagger \hat{a}_{-\kk'+\pp \downarrow} \hat{a}_{\kk' \uparrow} \right)
\end{multline}
where all the combinations of wave vectors have to be mapped back into the first Brillouin zone. 
To linearize these equations of motion, we consider a small region of the Hilbert space around the BCS ground state in which the action of the operators is only slightly different from {multiplication by} their BCS ground state expectation value noted 
$\meanv{\ldots}_0$. These average values will then be taken as zeroth order quantities (note that only the operators with $\qq=\mathbf{0}$ have a non-zero expectation value) from which the operators differ by a first order infinitesimal quantity. This suggest to write an arbitrary quadratic operator $\hat{a} \hat{b}$ {(where $\hat{a}$ and $\hat{b}$ are creation or annihilation operators)} as
\begin{equation}
\label{eq:expansionRPA}
\hat{a} \hat{b} = \langle \hat{a} \hat{b}\rangle_{0} + \delta(\hat{a} \hat{b})
\end{equation}
This prescription however is not sufficient. Indeed, a quartic operator $\hat{a} \hat{b} \hat{c} \hat{d}$ can be reordered using anticommutation rules and one cannot pair the operators inserting the first order expansion (\ref{eq:expansionRPA}) in a unique way. Instead, the RPA considers that a product $\hat{a} \hat{b} \hat{c} \hat{d}$ is of relevant order if one can form a  $\qq=0$ quadratic operator from at least two of the linear operators. Otherwise, the product will be regarded as second order and discarded. This procedure is equivalent to replacing the product using incomplete Wick's contractions:
{
\begin{multline}
\hat{a} \hat{b} \hat{c} \hat{d} \rightarrow \hat{a}\hat{b} \langle\hat{c}\hat{d} \rangle_0  + \langle\hat{a}\hat{b} \rangle_0\hat{c}\hat{d} -\hat{a}\hat{c} \langle\hat{b}\hat{d}\rangle_0 -  \langle\hat{a}\hat{c}\rangle_0 \hat{b}\hat{d} + \hat{a}\hat{d} \langle\hat{b}\hat{c}\rangle_0 \\ +  \langle\hat{a}\hat{d}\rangle_0\hat{b}\hat{c}   -  \langle\hat{a}\hat{b} \rangle_0  \langle\hat{c}\hat{d} \rangle_0 + \langle\hat{a}\hat{c} \rangle_0  \langle\hat{b}\hat{d} \rangle_0- \langle\hat{a}\hat{d} \rangle_0  \langle\hat{b}\hat{c} \rangle_0
\end{multline}
}
Note that the last three terms are included to ensure that the expectation value  $\langle \hat{a} \hat{b} \hat{c} \hat{d}\rangle_0$ remains exact 
in this approximation.
The simplification introduced by the RPA \textit{decouples} the operators of different $\qq$ so that we are left with a set of linear differential equations for each value of $\qq$. Furthermore the phase dynamics we are interested in takes place
in the $\qq=\mathbf{0}$ subspace where lives the anomalous mode due to the $U(1)$ symmetry breaking, a subspace 
to which we restrict by now. We have then from \eqref{eq:quadratic_operators} the simplifications
$\hat{n}_{\kk\uparrow}^{\mathbf{0}}=\hat{{n}}_{\kk\uparrow}$,
$\hat{\bar{n}}_{\kk\downarrow}^{\mathbf{0}}=\hat{{n}}_{-\kk\downarrow}$,
and
$\hat{\bar{d}}_\kk^{\mathbf{0}}=\hat{d}_\kk^{\mathbf{0} \dagger}$. As a shorthand notation we use
\be
\hat{{d}}_\kk \equiv \hat{{d}}_\kk^{\mathbf{0}}= \hat{a}_{-\kk \downarrow} \hat{a}_{\kk \uparrow} .
\ee
We also introduce as a more convenient combination of zero-mean variables: 
\begin{eqnarray}
\hat{y}_\kk=&\delta \hat{d}_\kk - \delta \hat{d}_\kk^\dagger \quad &\hat{s}_\kk=\delta \hat{d}_\kk + \delta \hat{d}_\kk^\dagger \\
\hat{m}_\kk=&\delta \hat{n}_{\kk\uparrow}+\delta \hat{n}_{-\kk\downarrow} \quad &\hat{h}_\kk=\delta \hat{n}_{\kk\uparrow}-\delta \hat{n}_{-\kk\downarrow}
\label{eq:variablesRPA}
\end{eqnarray}
where the $\delta$ indicates the deviation of the operator with respect to its expectation value in the BCS ground state:
$\delta \hat{x} \equiv \hat{x}-\langle \hat{x}\rangle_0$.
From the linear equations of motion (not given here) we remark that two linear combinations of these four variables are in fact constants of motion:
\begin{eqnarray}
&&\frac{d\hat{h}_{\kk}}{dt}=0 \label{eq:h_dot}\\
&&\frac{d \hat{\zeta}_\kk}{dt}=0 \quad \mbox{where} \quad \hat{\zeta}_\kk=\hat{s}_\kk+\frac{\xi_\kk}{\Delta_0}\hat{m}_{\kk}
\label{eq:z_dot}
\end{eqnarray}
The quantity $\hat{h}_\kk$ is indeed conserved when one creates or annihilates pairs of particles with opposite spin
and zero total momentum.
Remarkably, {the hermitian operator} $\hat{\zeta}_\kk$ has a zero mean and a zero variance in the BCS ground state:
\be
\hat{\zeta}_\kk |\psi_{\rm BCS}^0\rangle = 0
\label{eq:zetak_psi0}
\ee
To derive \eqref{eq:zetak_psi0} we expressed the various quantities in terms of $V_\kk^0$, {keeping in mind
the relation \eqref{eq:utile3}} :
\bea
\langle\hat{d}_\kk\rangle_0&=&  -\frac{\Delta_0}{2\epsilon_\kk} = -U_\kk^0 V_\kk^0
\label{eq:meanb} \\
2 \langle\hat{n}_{\kk\uparrow}\rangle_0&=& 1 - \frac{\xi_\kk}{\epsilon_\kk}= 2 (V_\kk^0)^2 
\label{eq:utile2}
\eea
We thus eliminate the redundant variable $\hat{s}_\kk$ in terms of $\hat{m}_\kk$ and $\hat{\zeta}_\kk$
to obtain the inhomogeneous linear system
\begin{equation}
\label{eq:RPA}
i \hbar \frac{d}{dt}
\begin{pmatrix} {\hat{y}_\kk}\\{ \hat{m}_{\kk}} \end{pmatrix}_{\kk \in {\cal D}}
= \mathcal{L}_{\rm RPA} \begin{pmatrix} {\hat{y}_\kk}\\{ \hat{m}_{\kk}} \end{pmatrix}_{\kk \in {\cal D}} + 
\begin{pmatrix} {\hat{S}_\kk}\\{ 0} \end{pmatrix}_{\kk \in {\cal D}}
\end{equation}
The source term is
\be
\hat{S}_\kk = 2 \xi_\kk \hat{\zeta}_\kk + \frac{g_0}{L^3} \frac{\xi_\kk}{\epsilon_\kk}\sum_{\qq} \hat{\zeta}_\qq
\label{eq:Skk}
\ee
Explicitly, the equations take the form
\begin{eqnarray}
\!\!\!\!\!\!\!-i\hbar\frac{d\hat{y}_{\kk}}{dt}&\!\!=\!\!&\frac{2\epsilon_\kk^2}{\Delta_0} \hat{m}_{\kk} + \frac{g_0}{L^3}\!\sum_\qq \!\bb{\frac{\xi_\kk \xi_\qq}{\epsilon_\kk \Delta_0}+\frac{\Delta_0}{\epsilon_\kk}} \hat{m}_{\qq}\! -\! \hat{S}_\kk
\label{eq:RPAexpli1}\\
\!\!\!\!\!\!\!-i\hbar\frac{d\hat{m}_{\kk}}{dt}&\!\!=\!\!&2\Delta_0 \hat{y}_{\kk} + \frac{g_0}{L^3} \frac{\Delta_0}{\epsilon_\kk}\sum_\qq  \hat{y}_{\qq} \label{eq:RPAexpli2}
\end{eqnarray}

Direct spectral decomposition of $\mathcal{L}_{\rm RPA}$ yields the zero energy mode
\be
\vec{e}_n^{\,\rm RPA}=\begin{pmatrix}2i\langle\hat{d}_\kk\rangle_0\\0 \end{pmatrix}_{\kk \in {\cal D}} 
 \mbox{and} \quad \mathcal{L}_{\rm RPA} \vec{e}_n^{\,\rm RPA} = \vec{0}
\label{eq:en_RPA}
\ee
where we used \eqref{eq:BCS2} and \eqref{eq:meanb}, 
and the anomalous mode
\be
\vec{e}_a^{\,\rm RPA}= \begin{pmatrix}0\\  2\frac{d}{d{\bar N}}\langle\hat{n}_{\kk \uparrow}\rangle_0 \end{pmatrix}_{\kk \in {\cal D}} \!\!\!\!\! \mbox{and} \ 
\mathcal{L}_{\rm RPA} \vec{e}_a^{\,\rm RPA} = -2i\frac{d\mu}{d\bar{N}}\vec{e}_n^{\,\rm RPA}
\label{eq:anomalous_RPA}
\ee
where we used the two intermediate relations
\bea
&& 2 \frac{d}{d{\bar N}} \langle \hat{n}_{\kk \uparrow}\rangle_0 =\frac{d\mu}{d{\bar N}} \frac{\Delta_0}{\epsilon_\kk^3}
\left[ \xi_\kk \frac{d\Delta_0}{d\mu} + \Delta_0 \bb{1-\frac{g_0}{2L^3}\frac{d{\bar N}}{d\mu}}\right] 
 \notag \\
&& \sum_\qq \xi_\qq \frac{d}{d{\bar N}}\langle\hat{n}_{\qq \uparrow}\rangle_0 =
- \frac{d\mu}{d{\bar N}}\frac{d\Delta_0}{d\mu} \frac{\Delta_0 L^3}{g_0} \quad 
\label{eq:interm2}
\eea
respectively obtained by taking the derivative of \eqref{eq:gap} and of the ground state version of \eqref{eq:Eclass},
with respect to ${\bar N}$ or ${\mu}$, and further using \eqref{eq:dNdmu} and the thermodynamic relation
$-{\bar N}= \frac{d E_0}{d\mu}$. Note that we normalized $\vec{e}_a^{\,\rm RPA}$ so that the second equality in \eqref{eq:anomalous_RPA} is identical to the one of the semiclassical theory \eqref{eq:anomalousmode}.

$\mathcal{L}_{\rm RPA}$ does not show the symplectic symmetry \eqref{eq:symplectique}, hence the necessity to perform the spectral analysis of  $\mathcal{L}_{\rm RPA}^\dagger$ to find the dual vectors defined as in
\eqref{eq:duals}. We obtain \cite{Nadal} for the dual of the zero-energy mode:
\be
\vec{d}_n^{\,\,\rm RPA}= \begin{pmatrix} {i (\frac{d}{d{\bar N}}\langle\hat{n}_{\kk\uparrow }\rangle_0)/\meanv{\hat{d}_\kk}_0}\\0 \end{pmatrix}_{\kk \in {\cal D}}
\label{eq:dnRPA}
\ee
and for the dual of the anomalous mode
\be 
\vec{d}_a^{\,\,\rm RPA} = \begin{pmatrix} 0 \\  1 \end{pmatrix}_{\kk \in {\cal D}}
\label{eq:daRPA}
\ee
{
Remarkably the matrix $\mathcal{L}_{\rm RPA}$ \eqref{eq:RPA} and the corresponding {Bogoliubov-de Gennes} matrix 
$\mathcal{L}$ \eqref{eq:timedependent} are related through a change of basis 
\begin{equation}
\mathcal{L}_{\rm RPA} = \Lambda \: \mathcal{L} \: \Lambda^{-1}  \label{eq:changement_base}
\end{equation}
}
and so are their modes 
\begin{equation}
\begin{pmatrix} { y_\kk }\\{ m_\kk } \end{pmatrix} = \Lambda_\kk \begin{pmatrix} {\delta V_\kk} \\{\delta V_\kk^{*}} \end{pmatrix} \qquad \begin{pmatrix} { y_\kk }\\{ m_\kk } \end{pmatrix}_d = \bb{\Lambda_\kk^{\dagger}}^{-1} \begin{pmatrix} {\delta V_\kk} \\{\delta V_\kk^{*}} \end{pmatrix}_d 
\end{equation}
where the subscript $d$ in $\bb{}_d$ refers to the dual vectors. Here $\Lambda_\kk$ is a two-by-two matrix
\begin{equation}
\Lambda_\kk=\begin{pmatrix}  -U_\kk^0  && U_\kk^0 \\ 2V_\kk^0 && 2V_\kk^0 \end{pmatrix}
\end{equation}
and $\Lambda$ is a block-diagonal matrix with matrices $\Lambda_\kk$, $\kk \in {\cal D}$ on the diagonal.
To show this correspondence, we think of the \textit{classical} fluctuations $\delta \langle \hat{d}_\kk \rangle = \langle \hat{d}_\kk \rangle - \langle \hat{d}_\kk \rangle_{0}$ and 
$\delta \langle \hat{n}_{\kk \uparrow} \rangle = \langle \hat{n}_{\kk \uparrow} \rangle - \langle \hat{n}_{\kk \uparrow} \rangle_{0}$
(where the expectation value $\meanv{\ldots}$ is taken in a BCS state of the form
\eqref{eq:usual_BCS} slightly perturbed away from the BCS ground state) as a particular case of the quantum fluctuations we consider here. In the state \eqref{eq:usual_BCS} $\langle 	\hat{d}_\kk \rangle=-U_\kk V_\kk$ according to
\eqref{eq:defDelta}, 
and $\langle \hat{n}_{\kk \uparrow} \rangle=V_\kk V_\kk^*$. Linearizing these relations around the BCS ground state one gets
\bea
\hat{y}_\kk &\leftrightarrow& - U_\kk^0 \bb{\delta V_\kk -\delta V_\kk^*} \label{eq:corr_y}\\
\hat{m}_\kk &\leftrightarrow& 2 V_\kk^0 \bb{\delta V_\kk +\delta V_\kk^*}  \label{eq:corr_m}
\eea
which explains the value of the matrix $\Lambda_\kk$.
One has also the correspondence $\hat{h}_\kk \leftrightarrow 0$ and $\hat{\zeta}_\kk \leftrightarrow 0$.
The equivalence \eqref{eq:changement_base} between the matrices is shown in Appendix \ref{app:equations_semi}.

In the RPA quantum theory, $\hat{P}$ and $\hat{Q}$ are now the amplitudes of the vector $\begin{pmatrix} { \hat{y}_\kk}&{ \hat{m}_\kk} \end{pmatrix}_{\kk \in {\cal D}}$ on the anomalous and zero energy mode respectively:
\be
\begin{pmatrix} { \hat{y}_\kk}\\{ \hat{m}_\kk} \end{pmatrix}_{\kk \in {\cal D}}=\hat{Q}\vec{e}_n^{\,\rm RPA} +
\hat{P}\vec{e}_a^{\,\rm RPA} {+ \ldots}
\label{eq:expansionfieldRPA}
\ee
{where we have written explicitly the expansion within the zero-energy subspace only.}
Using the expression of the dual vectors \eqref{eq:dnRPA} and  \eqref{eq:daRPA} we obtain:
\bea
\hat{P} &=&  \sum_\kk  \hat{m}_\kk = \delta \hat{N} \\
 \hat{Q}&=& 2 i  \frac{d\mu}{d\bar{N}} \sum_\kk \frac{ \frac{d}{d\mu} V_\kk^0}{U_\kk^0} 
 \bb{ \delta \hat{d}_\kk- \delta \hat{d}_\kk^\dagger}
 \label{eq:Q_RPA}
\eea
The resulting equations of motion are
\bea
\frac{d\hat{P}}{dt}&=&0 \label{eq:hatPdot}\\
\frac{d\hat{Q}}{dt}&=&-\frac{2}{\hbar}\frac{d\mu}{d\bar{N}}  \hat{P} + \hat{Z}\,. \label{eq:hatQdot}
\eea
The difference with the corresponding {linearized time-dependent Bogoliubov-de Gennes} equations \eqref{eq:Pdot} and \eqref{eq:Qdot} is due to the 
operator $\hat{Z}=(2/\hbar)\sum_\kk \frac{d V_\kk^0}{d {\bar N}} \hat{S}_\kk/U_\kk^0$ that originates 
from the source term in \eqref{eq:RPAexpli1}. Its mean and variance in the BCS ground state are however zero,
\be
\hat{Z} |\psi_{\rm BCS}^0\rangle=0
\label{eq:Z0}
\ee
so that this operator will not contribute to the collapse of the order parameter in the thermodynamic limit
as we will see.


\subsection{Collapse of the order parameter}


\label{sec:collapse_eiQfactor}
We can now calculate the evolution of the order parameter for a system initially prepared in the BCS ground state. In the Heisenberg picture, 
\be
\Delta(t) {\equiv g_0 \meanv{\hat{\psi}_\downarrow(t) \hat{\psi}_\uparrow(t)}_{0} }= 
\frac{g_0}{L^3}\sum_\kk \meanv{ \hat{d}_\kk(t)}_{0} 
\ee
From \eqref{eq:expansionfieldRPA} and using \eqref{eq:Uk},\eqref{eq:utile3},\eqref{eq:meanb},\eqref{eq:utile2}, we obtain the evolution of $\hat{d}_\kk$ in the zero energy subspace:
\be
\hat{d}_\kk(t)=[1+i\hat{Q}(t)]\meanv{\hat{d}_\kk}_{0}+\delta \hat{N} \frac{d}{d{\bar N}} \meanv{\hat{d}_\kk}_{0} 
+\frac{1}{2} \hat{\zeta}_\kk(0) {+ \ldots} \label{eq:hatb}
\ee
Should we calculate the order parameter directly from this expression we would {obtain} a constant value because all the operators arising from the linearized equations of motion have a zero mean value. To overcome this difficulty, we 
recall that, for an arbitrary (i.e not necessarily infinitesimal) phase fluctuation {$Q$}, the field is modified as
\be
V_\kk=e^{i Q}V_\kk^0 \qquad {\meanv{\hat{d}_\kk}}=e^{i Q}\meanv{\hat{d}_\kk}_0
\ee
The $1+i\hat{Q}$ terms appearing in our decomposition must therefore be the linearization of the operator $e^{i\hat{Q}}$ contained in $\hat{d}_\kk(t)$. Having recovered this factor, the order parameter reads:
\be
\Delta(t)=\Delta_0 \meanvlr{e^{i\hat{Q}(t)}}_{0}
\ee
Using $\hat{Q}(t)=\hat{Q}(0) {-} 2 \frac{d\mu}{d\bar{N}}  (\hat{N}-\bar{N})\frac{t}{\hbar}$ and going to the thermodynamic limit where {one can neglect the contributions of $\hat{Z}$ and of the fluctuations of the initial phase $\hat{Q}(0)$}, we obtain
\be
\Delta(t)=\Delta_0 e^{-2t^2/t_{\rm br}^2}
\label{eq:Delta_t}
\ee
where the phase blurring time $t_{\rm br}$ is given by \eqref{eq:tbr}.
We give the details of this calculation in Appendix \ref{sec:appendixcalculcollapse}.


\subsection{Application: blurring time in the BEC-BCS crossover}


The mean field BCS theory gives analytical expressions of the thermodynamical quantities {in the BEC-BCS} crossover 
{\cite{Leggett1980,Engelbrecht1997,RanderiaGreenBook}} {in terms of special functions} \cite{Strinati1998}. {We use them here to
show} some numerical results {for} the coherence time of a BCS ground state, paying a special attention to the so-called BEC and BCS limits.

\begin{figure}[!h]
\includegraphics[width=0.5\textwidth]{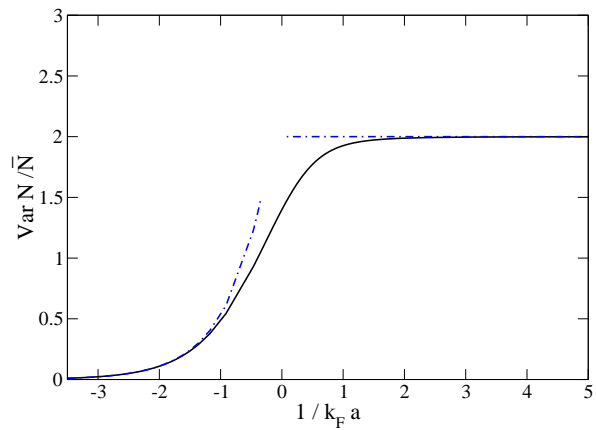}
\caption{(Color online) Variance of {the total particle number} $\hat{N}$ in the BCS ground state (solid line). The dashed-dotted blue lines are the asymptotic behaviors in the BEC or BCS limits. 
\label{fig:variance}}
\end{figure}

We imagine that we have initially prepared a BCS ground state so that the variance of the particle number 
can be expressed as a sum over $\kk$ using {\eqref{eq:meanb}}:
\begin{equation}
\text{Var\,}\hat{N}= \sum_\kk \frac{\Delta_0^2}{\epsilon_\kk^2}
\label{eq:varianceN}
\end{equation} 
$\text{Var\,}\hat{N}$ is an explicit function of $\mu$ and $\Delta_0$. Using BCS equations (\ref{eq:BCS1}) and (\ref{eq:BCS2}) we express it in terms of {the Fermi wavenumber $k_F=(6\pi^2 \rho_\sigma^{0})^{1/3}$} and of the scattering length $a$ as we change the sum in \eqref{eq:varianceN} into an integral in the thermodynamic limit  and as we take the limit of a vanishing lattice step $b$. The result is shown in Fig.~\ref{fig:variance}.
In the $k_F a \rightarrow 0^-$ BCS limit, the variance is proportional to the gap and thus tends exponentially to zero:
\begin{equation}
\text{Var\,}\hat{N} \underset{k_F a \rightarrow 0^-}{\sim} \frac{3 \pi}{4} \frac{\bar{N}}{\epsilon_F} \Delta_0 = \frac{3 \pi\bar{N}}{4} {8 e^{-2}} e^{-\frac{\pi}{2 k_F |a|}}
\end{equation} 
{where $\epsilon_F=\hbar^2 k_F^2/(2m)$ is the Fermi energy.}
In the  $k_F a \rightarrow 0^+$ BEC limit, the BCS theory correctly predicts the Poissonian variance of an ideal gas of composite bosons $\text{Var\,}(\hat{N}/2)=\bar{N}/2$.

\begin{figure}[!h]
\includegraphics[width=0.5\textwidth]{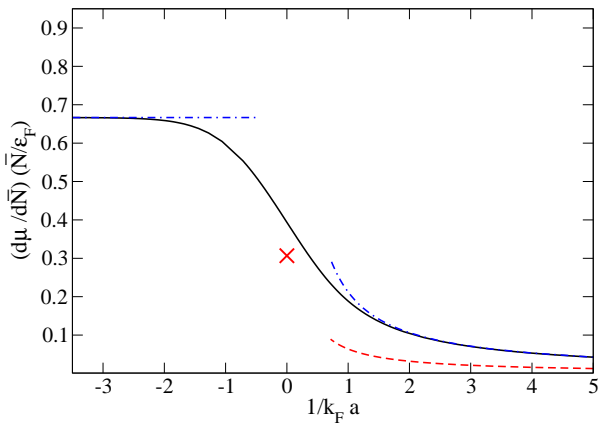}
\caption{(Color online) $d \mu/d\bar{N}$ in the BCS ground state. The full line is the BCS mean field approximation, the dashed-dotted blue lines show the asymptotic behaviors. The dashed red line is the exact (beyond BCS theory) asymptotic behavior on the BEC side and the red cross is the value at the unitary limit $k_F |a|= \infty$
deduced from the value of the universal parameter $\xi$ in the equation of state \cite{Zwerger2012}.
\label{fig:dmudN}}
\end{figure}

{The derivative} $d \mu/d\bar{N}$ is the variation with ${\bar N}$ of the energy cost of adding an extra particle to the gas when $\bar{N}$ in average are already present. For a BCS state, it can be obtained by taking the derivative of the mean density equation (\ref{eq:BCS1}) with respect to $\mu$.  {When $b\to0$, the coupling constant $g_0\to0$ and}
\begin{equation}
\frac{d \mu}{d\bar{N}}= \frac{\Delta_0 \Theta}{\Theta^2 + X^2}
\label{eq:dmudN_XandTheta}
\end{equation} 
where we have defined the quantities
\begin{equation}
\Theta =  \sum_\kk \frac{\Delta_0^3}{\epsilon_\kk^3} \qquad X= \sum_\kk \frac{\Delta_0^2 \xi_\kk }{\epsilon_\kk^3}
\end{equation}
{related to $\Delta_0$ by
\be
\frac{d\Delta_0}{d\mu}=\frac{X}{\Theta}
\label{eq:dDelta_dmu}
\ee}
{as it can be obtained by taking the derivative of \eqref{eq:BCS2} with respect to $\mu$.}
The behavior of {$d\mu/d\bar{N}$} is shown in Fig. \ref{fig:dmudN} (see also \cite{Belzig2007}).
Since adding an extra fermion to a Fermi sea costs an energy $\epsilon_F$, $\mu$ should tend to $\epsilon_F$ and $d \mu/d\bar{N}$ to $d\epsilon_F/d\bar{N}=(2/3) \epsilon_F/\bar{N}$ when $k_F a \rightarrow0^-$, as is correctly reproduced by the BCS approximation. In the BEC limit, the approximation predicts for $\mu$
\begin{equation}
\mu \underset{k_F a \rightarrow 0^+}{=} - \frac{\epsilon_F}{(k_F a)^2} + \frac{\epsilon_F }{3 \pi} k_F a_{\rm mol} + o_{}(k_F a)
\end{equation}
which is the mean field chemical potential of a gas of dimers with binding energy $\hbar^2/ ma^2$ and dimer-dimer scattering length $a_{\rm mol}$. Then for $d \mu/d\bar{N}$
\begin{equation}
d \mu/d\bar{N} \underset{k_F a \rightarrow 0^+}{=}  \frac{\epsilon_F }{3 \pi \bar{N}} {k_F a_{\rm mol}} + o_{}(k_F a)
\end{equation}
The value of the dimer-dimer scattering length predicted by the BCS theory $a_{\rm mol}=2a$ {\cite{Pieri2000}} is however quantitatively incorrect. The exact value $a_{\rm mol}=0.6a$ obtained in reference \cite{Petrov2004,Leyronas2006} is  used to plot the dashed red line of Fig.~\ref{fig:dmudN}.

\begin{figure}[h]
\includegraphics[width=0.5\textwidth]{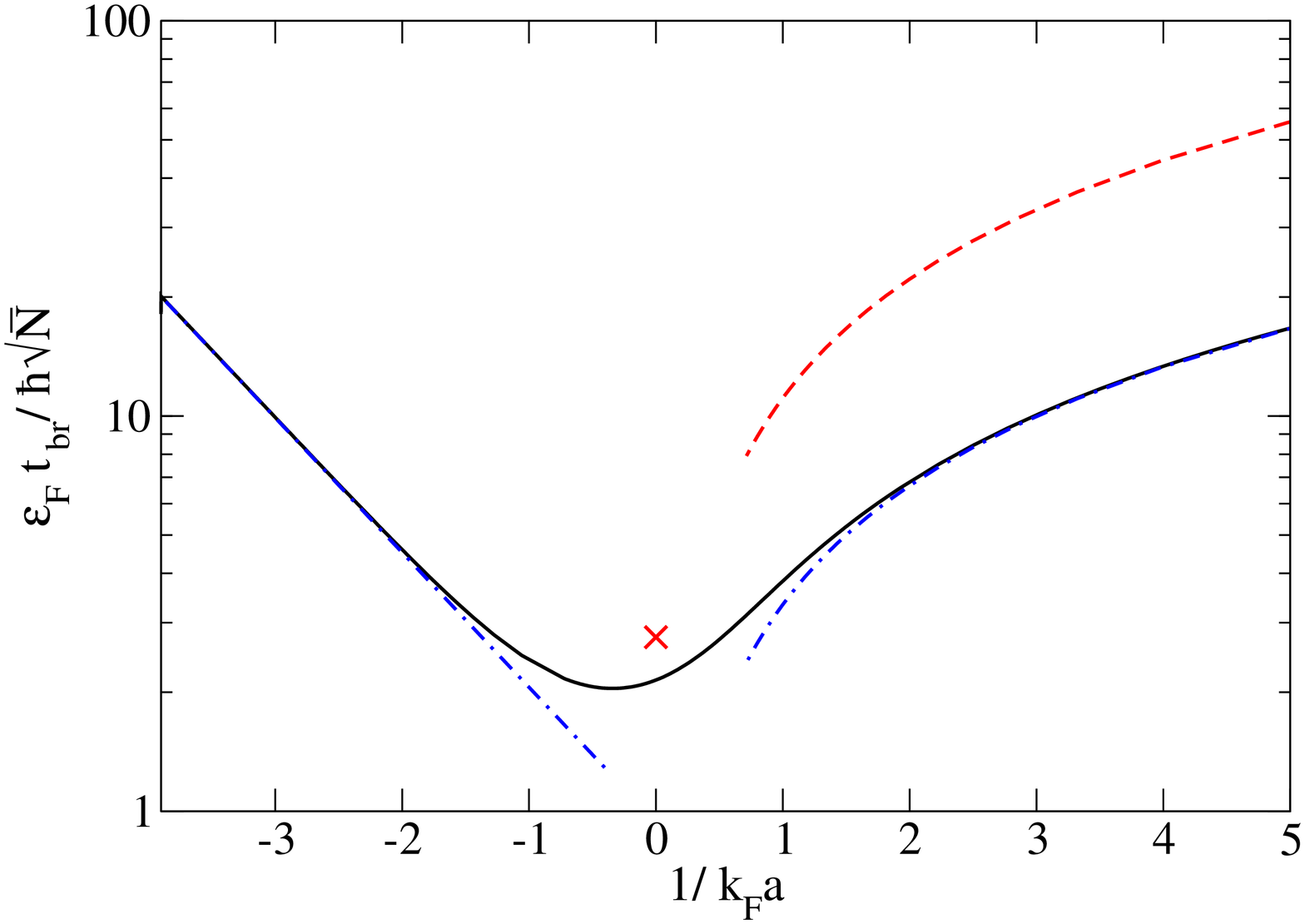}
\caption{(Color online) Phase blurring time $t_{\rm br}$ in the BCS ground state (solid line). The dashed-dotted blue lines show the asymptotic behaviors. The dashed red line is the exact asymptotic behavior and the red cross is the expected value for $k_F |a|= \infty$. In all cases we take $\mbox{Var \,} \hat{N}$ as given by the BCS ground state.
\label{fig:tbr}}
\end{figure}

The blurring time is shown in Fig.~\ref{fig:tbr}. 
It tends to infinity in both BEC and BCS limits, however not for the same reasons. In the BEC limit, as it is the case in bosonic phase dynamics, the blurring time diverges because the non-linearity of the Hamiltonian introduced by the $d \mu/d\bar{N}$ factor vanishes:
{
\begin{equation}
t_{\rm br}/\hbar \underset{k_F a \rightarrow 0^+}{\sim} \frac{3 \pi \sqrt{\bar{N}}}{\sqrt{2}{\epsilon_F} k_F a_{\rm mol}}
\end{equation}
}
Again the exact value $a_{\rm mol}=0.6a$ is used to plot the dashed red line.
In the BCS limit however, the divergence is due the fact that the initial variance $\text{Var\,} \hat{N}$ tends to zero as the BCS ansatz converges to the Fermi sea of the ideal gas:
{
\begin{equation}
t_{\rm br}/\hbar \underset{k_F a \rightarrow 0^-}{\sim}  \left( \frac{3e^2}{8\pi} \right)^{1/2}  \frac{\sqrt{\bar{N}}}{\epsilon_F} 
e^{+\frac{\pi}{4 k_F |a|}}
\end{equation}
}
{In the whole interaction range, the blurring time is proportional to $\sqrt{\bar{N}}$ and diverges in the thermodynamic limit as expected.} 
{We emphasize however} that the particle number variance of the $U(1)$ symmetry breaking BCS state 
does not have in fact any physical meaning \cite{Combescot2007}. 
We use it here as an illustration of our results. In practice the 
variance of $\hat{N}$, {and hence the $N$ dependence of the blurring time,} will depend on the details of the realization of the interference experiment (see 
Sec.~\ref{sec:hintsforanexperiment}).


\section{What the adiabatic phase operator really is}


\label{sec:phases}

Surprisingly, the phase operator $\hat{Q}$ that appears in our dynamical study \textit{is not} the phase of the condensate of pairs, $\hat{\theta}_0$, that we introduce in this section. 


\subsubsection{Phase of the condensate}


To define the phase operator of the condensate, we assume that the state of the gas is such that one and only one mode, noted $\phi(\rr_1,\rr_2)$, of the two body density matrix is macroscopically populated. This mode is defined by the eigenvalue problem (see section 2.4 in \cite{Leggett2006})
\be
b^6 \sum_{\rr_1',\rr_2'}    \rho_2(\rr_1,\rr_2 ; \rr_1',\rr_2') \phi(\rr_1',\rr_2')= {\bar N}_0  \phi(\rr_1,\rr_2)
\label{eq:densitymatrixeigenvalue}
 \ee
 where $\rho_2$ is the opposite spin two-body density matrix in real space 
 \be
 \rho_2(\rr_1,\rr_2 ; \rr_1',\rr_2')=\meanv{\hat{\psi}_\uparrow^\dagger({\rr_1'}) \hat{\psi}_\downarrow^\dagger(\rr_2')\hat{\psi}_\downarrow(\rr_2)\hat{\psi}_\uparrow(\rr_1)} \,.
 \ee
 Here $\phi$ is the \textit{condensate wave-function} and ${\bar N}_0 $, which scales as $N$, is the \textit{number of condensed pairs}.
 
This is indeed the case of the BCS ground state. The density matrix computed using Wick's contractions contains two non zero terms:
\begin{multline}
 \rho_2(\rr_1,\rr_2 ; \rr_1',\rr_2')=\meanv{\hat{\psi}_\uparrow^\dagger\bb{\rr_1'} \hat{\psi}_\downarrow^\dagger\bb{\rr_2'}}_0\meanv{\hat{\psi}_\downarrow\bb{\rr_2}\hat{\psi}_\uparrow\bb{\rr_1}}_0 \\ +\meanv{\hat{\psi}_\uparrow^\dagger\bb{\rr_1'} \hat{\psi}_\uparrow\bb{\rr_1}}_0 \meanv{ \hat{\psi}_\downarrow^\dagger\bb{\rr_2'} \hat{\psi}_\downarrow\bb{\rr_2}}_0 \,.
\label{eq:rho2_factorized}
\end{multline}
The second term involves functions of $\rr_1-\rr_1'$ and $\rr_2-\rr_2'$ which tend to zero on typical scales given by
the Fermi length $k_F^{-1}$ or the pair size $\hbar^2 k_F/m \Delta_0$. 
In the thermodynamic limit, 
keeping only the long range (LR) part, we obtain a factorized density matrix:
\begin{multline}
\rho_2^{\text{LR}}(\rr_1,\rr_2 ; \rr_1',\rr_2')= \frac{1}{L^6}\sum_{\kk,\kk'} \frac{\Delta_0^2}{4\epsilon_{\kk} \epsilon_{\kk'}}  e^{-i \kk \cdot (\rr_1-\rr_2)} e^{i \kk' \cdot (\rr_1'-\rr_2')} \\
 ={\bar N}_0 \phi(\rr_1,\rr_2)\phi^*(\rr_1',\rr_2')\,.
\end{multline}
The only populated eigenvector of this matrix, here normalized to unity, does not depends on the center of mass coordinates:
\be
\phi(\rr_1,\rr_2)= \frac{1}{\sqrt{{\bar N}_0 }L^3} \sum_{\kk} \frac{\Delta_0}{2 \epsilon_\kk} e^{-i\kk \cdot (\rr_1-\rr_2)}
\ee
Even though we deal with an homogeneous system, this  wave function depends (via $\epsilon_\kk$) on the total number of particles in the system.
The corresponding number of condensed particles is
\be
{\bar N}_0 = \sum_\kk \frac{\Delta_0^2}{4 \epsilon_\kk^2}
\ee
Remarkably the condensed fraction $2 {\bar N}_0 /{\bar N}$ is equal to the quantity $\text{Var\,}\hat{N}/2\bar{N}$ already shown in Fig.~\ref{fig:variance}, as it is apparent from equation \eqref{eq:varianceN}. In the ground state ${\bar N}_0 /\bar{N}$ has a fixed value for a given interaction strength $k_F a$.  Changing this ratio by adding new particles in the condensate will excite the system. In the BEC limit all the composite bosons are condensed, whereas in the BCS limit the number of condensed Cooper pairs goes to zero as $\Delta_0\to0$ and the state of the system approaches the Fermi sea. 

 The amplitude of the field of pairs on the condensate mode is obtained by projection onto the condensate wave-function:
 \be
 \label{eq:amplitudecondensate}
 \hat{a}_0 \equiv b^6 \sum_{\rr_1  \rr_2} \phi^*(\rr_1,\rr_2)  \hat{\psi}_\downarrow(\rr_2) \hat{\psi}_\uparrow(\rr_1)
 \ee
And the \textit{phase of the condensate} is the phase of this amplitude
 \be
  \label{eq:phasecondensate}
e^{-i\hat{\theta}_0}=\frac{1}{\sqrt{\hat{N}_0}} \hat{a}_0^\dagger
 \ee
 where $\hat{N}_0=\hat{a}_0^\dagger \hat{a}_0$ {and $\hat{\theta}_0$ is hermitian.}
In the BEC limit,  $\hat{a}_0$ obeys bosonic commutation relations and {the phase operator is well defined if we neglect border effects for an empty condensate mode \cite{Nieto1968}. It then} generates translations of the number of condensed dimers:
 \be
\meanv{e^{i \delta N{_0} \hat{\theta}_0} \hat{a}_0^\dagger \hat{a}_0 e^{-i \delta N{_0} \hat{\theta}_0}}={\bar N}_0 +\delta N{_0} 
\label{eq:translationBEC}
 \ee
Out of the BEC limit, $\hat{a}_0$ is not a bosonic operator and the translation \eqref{eq:translationBEC} does not hold \endnote{This last remark raises the question of the validity of equation \eqref{eq:phasecondensate}. From the alleged unitarity of $e^{-i\hat{\theta}_0}$, one expects
$$
\hat{a}_0 \frac{1}{\hat{N}_0}  \hat{a}_0^\dagger= \frac{1}{\sqrt{\hat{N}_0}} \hat{a}_0^\dagger \hat{a}_0 \frac{1}{\sqrt{\hat{N}_0}} =1
$$
While the second equality is obvious, to obtain the first one, one can use the relation
$$
\bb{1+ \frac{1}{\hat{N}_0}[\hat{a}_0,\hat{a}_0^\dagger]}\hat{a}_0\frac{1}{\hat{N}_0}\hat{a}_0^\dagger=\bb{1+ \frac{1}{\hat{N}_0}[\hat{a}_0,\hat{a}_0^\dagger]}
$$
that can be checked by straightforward expansion of the commutators. In a small neighborhood of the BCS ground state, one can show that $[\hat{a}_0,\hat{a}_0^\dagger]=O(1)$, so that in the large $N_0$ limit the operator in parenthesis is invertible and can be simplified.}.
In the BCS approximation the phase  can be expressed analytically:
\be
e^{-i\hat{\theta}_0}=\frac{1}{2\sqrt{{\bar N}_0 \hat{N}_0}} \sum_{\kk} \frac{\Delta_0}{ \epsilon_\kk} \hat{a}_{{\kk} \uparrow}^\dagger \hat{a}_{-{\kk} \downarrow}^\dagger
\label{eq:phaseofthecond_exp}
\ee
If we linearize \eqref{eq:phaseofthecond_exp}, close to the BCS ground state, we obtain an expression with a structure similar to \eqref{eq:Q_RPA}:
 \be
\hat{\theta}_0=\frac{i}{4{\bar N}_0} \sum_{\kk} \frac{\Delta_0}{ \epsilon_\kk}\bb{ \delta \hat{d}_{{\kk}}- \delta \hat{d}_{{\kk}}^\dagger } \label{eq:phaseofthecond}
 \ee
but the coefficients on each mode of wave vector $\kk$ are different so that $\hat{\theta}_0$ is not in general equal to $\hat{Q}$. 

\begin{figure}[h!]
\includegraphics[width=0.5\textwidth]{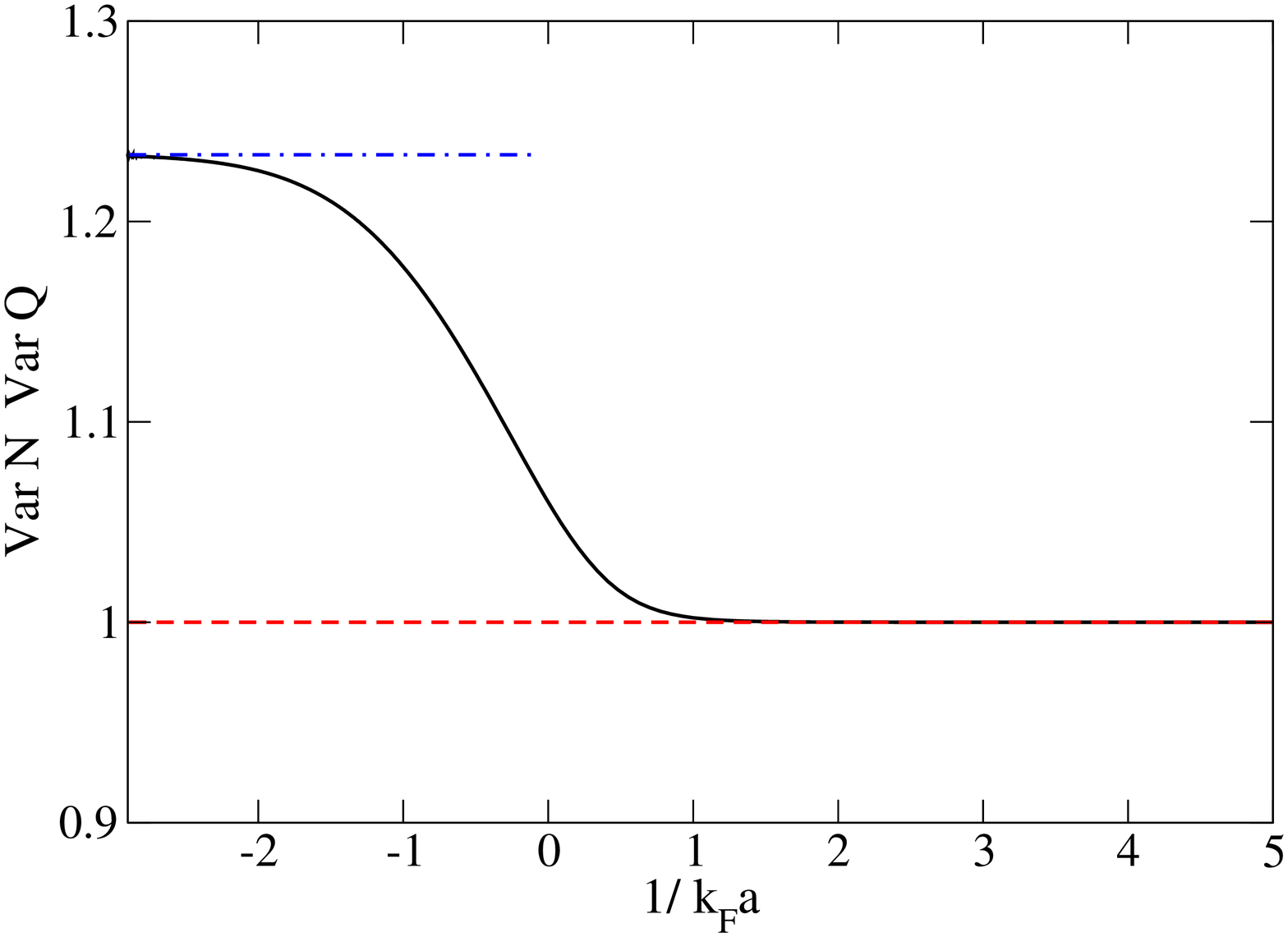}
\caption{(Color online) Product of the variances of the phases with the variance of the number of particles at time $t=0$
{for a system in the BCS ground state}. The black solid line shows $ \text{Var\,}\hat{Q}{(0)}  \text{Var\,}\hat{N}$, the dashed red line $\text{Var\,}\hat{\theta}_0{(0)}  \text{Var\,}\hat{N}$, and the blue dashed dotted line is the BCS limit of  $ \text{Var\,}\hat{Q}{(0)}  \text{Var\,}\hat{N}$. {The expression of $\text{Var\,}\hat{Q}{(0)}$ is given {
by \eqref{eq:varQ0}} and {the one of} $\text{Var\,}\hat{N}$ by equation \eqref{eq:varianceN}.}
\label{fig:produitvariances}}
\end{figure}

We note that the variance of $\hat{\theta}_0$ (at time $t=0$) has the property
\be
\text{Var\,}\hat{\theta}_0(0)  \text{Var\,}\hat{N}=1
\label{eq:varthetavarN}
\ee
In Fig. \ref{fig:produitvariances} we show, as a function of $1/k_F a$, $ \text{Var\,}\hat{Q}(0)  \text{Var\,}\hat{N}$ and $\text{Var\,}\hat{\theta}_0(0)  \text{Var\,}\hat{N}$, the latter being identically equal to one \eqref{eq:varthetavarN} and the former being larger than one. The value of $\text{Var\,}\hat{Q}(0)$ is given in \eqref{eq:varQ0}. The variances of the two phases coincide only in the BEC limit. 


\subsubsection{The adiabatic phase shifts the number of particles in the ground state}


To explain the difference between $\hat{Q}$ and $\hat{\theta}_0$, we study the action of $e^{i\hat{Q}}$ on the BCS ground state.
Using the expression \eqref{eq:Q_RPA} we write 
\begin{multline}
e^{i\delta N \hat{Q}}=\prod_\kk \exp\bb{ {-} \frac{2\delta N\frac{d}{d{\bar N}}V_\kk^0}{U_\kk^0} \hat{y}_\kk} \\ \underset{\delta N\ll \bar{N}}{\simeq} \prod_\kk \bb{1 {-} \frac{2\delta N\frac{d}{d{\bar N}}V_\kk^0}{U_\kk^0}\hat{y}_\kk} 
\end{multline}
Then acting on the BCS ground state we obtain
\begin{multline}
\prod_\kk \bb{1  {-} \frac{2\frac{d}{d{\bar N}}V_\kk^0}{U_\kk^0}\hat{y}_\kk} \ket{\psi_{\rm BCS}^0}\\=\prod_\kk \left[ \bb{U_\kk^0 {-} 2\delta N \frac{d}{d{\bar N}}U_\kk^0} \right.\\ \left. -\bb{V_\kk {-} 2\delta N \frac{d}{d{\bar N}}V_\kk^0}\hat{a}_{{\kk} \uparrow}^\dagger \hat{a}_{-{\kk} \downarrow}^\dagger \right] \ket{0} 
\end{multline}
and hence the property
\begin{equation}
e^{i\delta N \hat{Q}} |\psi_{\rm BCS}^0(\bar{N})\rangle=|\psi_{\rm BCS}^0(\bar{N} {-} 2\delta N)\rangle
\end{equation}
$\hat{Q}$ is then the generator of \textit{adiabatic} translations of the number of pairs in the ground state.
In the BEC limit Pauli blocking plays no role and the ground state is a pure condensate of dimers, ${\bar N}_0=\bar{N}/2$. We are then not surprised that translating the ground state or translating the number of condensed particles be strictly equivalent:
\be
\hat{Q}\underset{k_F a \rightarrow 0^+}{\to}\hat{\theta}_0
\ee
as we have checked explicitly. 
In other regimes, Pauli blocking cannot be neglected. Adding a new Cooper pair in the ground state thus requires to update the wave function $\phi$. $\hat{Q}$ does this updating whereas $\hat{\theta}_0$ creates excitations out of the BCS ground state.

Using the RPA equation \eqref{eq:RPAexpli1} we can evaluate the time derivative of $\hat{\theta}_0$. For simplicity
we give the results only in the zero lattice spacing limit {(more details are given App. \ref{app:theta_0}):}
\be
\frac{d\hat{\theta}_0}{dt} \underset{b\to0}{=}-\frac{1}{2 \hbar {\bar N}_0}\sum_\kk \left[ \hat{m}_\kk (\epsilon_\kk -\xi_\kk) - 
\hat{S}_\kk \frac{\Delta_0}{2 \epsilon_\kk} \right]
\label{eq:dtheta}
\ee
where the contribution involving $\hat{m}_\kk$ is not proportional to $\hat{P}$ except in the BEC limit. $\hat{\theta}_0$ thus has a projection on the excited modes of the RPA equations. 
{However we argue that the long time dynamics of the two phases is the same. We expand
$\hat{y}_\kk$ over the eigenmodes of ${\cal L}$ using \eqref{eq:hatb} including 
the projection $\hat{y}_\kk^{\rm exc}$ of $\hat{y}_\kk$ over the excited modes of ${\cal L}$:
\be
\hat{y}_\kk = 2i \hat{Q} \meanv{\hat{d}_\kk}_0 + \hat{y}_\kk^{\rm exc}
\ee
Inserting this expression in \eqref{eq:phaseofthecond} and using \eqref{eq:meanb}, we find remarkably
\be
\hat{\theta}_0 = \hat{Q} + \frac{i}{4 \bar{N}_0} \sum_\kk \frac{\Delta_0}{\epsilon_\kk} \hat{y}_\kk^{\rm exc}
\label{eq:theta0_Q_plus}
\ee
At long times, the second term on the right hand side of \eqref{eq:theta0_Q_plus}, which is a sum of oscillating terms, is negligible with respect to the first term that is linear in time according to \eqref{eq:hatQdot}.}


\subsubsection{The bosonic case revisited}


We show here that even for bosons, the adiabatic phase and the phase of the condensate do not coincide if the
condensate wave function depends on the particle number as for example in the trapped case. 


\paragraph{Eigenmodes of the linearized equations}


The adiabatic phase $\hat{Q}^B$ naturally appears in the dynamics when the number of particles is not fixed.  
To set a frame, we consider the symmetry breaking description in section V of \cite{CastinDum1998}.  
The linearized dynamics is ruled by the operator ${\cal L}_{\rm GP}$ obtained by linearization of the Gross-Pitaevskii equation:
\be
i \hbar \frac{d}{dt}  \begin{pmatrix} {\delta \hat{\psi}} \\ {\delta \hat{\psi}^\dagger} \end{pmatrix}  = {\cal L}_{\rm GP}  
 \begin{pmatrix} {\delta \hat{\psi}} \\ {\delta \hat{\psi}^\dagger} \end{pmatrix}
 \label{eq:ldB}
\ee
The field fluctuations of the bosonic field operator are $\delta \hat{\psi}=\hat{\psi}-\Psi_0$ where we introduced
the ground state order parameter $\Psi_0=\meanv{\hat{\psi}}_0$ taken to be the positive solution of the Gross-Pitaevskii equation
\be
(h_0 -\mu + g_0 \Psi_0^2) \Psi_0 = 0\,,
\label{eq:GPE}
\ee
and 
\be
{\cal L}_{\rm GP} = \begin{pmatrix} h_0 + 2 g_0 \Psi_0^2 -\mu & g_0 \Psi_0^2  \\
-g_0 \Psi_0^2  & -(h_0 + 2 g_0 \Psi_0^2 -\mu)\end{pmatrix} \,.
\ee
Here $h_0$ is the one-body Hamiltonian including kinetic energy and trapping potential.
In addition to the usual Bogoliubov modes, of {positive} energy {$\epsilon_\lambda^B$}
{and modal functions $U_\lambda^{B}(\rr)$ and $V_\lambda^{B}(\rr)$}, such that
\be
{\cal L}_{\rm GP} \begin{pmatrix} {U_\lambda^{B}} \\ {V_\lambda^{B}} \end{pmatrix}  = \epsilon_\lambda^B 
 \begin{pmatrix} {U_\lambda^{B}} \\ {V_\lambda^{B}} \end{pmatrix}
 \label{eq:UkVk}
\ee
{with $\sum_\rr b^3 [|U_\lambda^{B}(\rr)|^2-|V_\lambda^{B}(\rr)|^2]=1$},
${\cal L}_{\rm GP}$ has a zero-energy and an anomalous mode.
With similar notations to those of section \ref{sec:collapse} \footnote{Note that to be consistent with out previous notations, the zero energy mode and the anomalous mode we introduce here differ from the ones of \cite{CastinDum1998} by a factor $\sqrt{\bar{N}}$ and $1/\sqrt{\bar{N}}$ respectively.}  
\bea
\vec{e}_n^{\,B} &=& \begin{pmatrix} {i \Psi_0} \\{-i \Psi_0} \end{pmatrix} 
\quad{\rm and} \quad
\vec{e}_a^{\,B} = 
\begin{pmatrix} {\frac{d}{d{\bar N}} \Psi_0} \\{\frac{d}{d{\bar N}} \Psi_0} \end{pmatrix} \\
\vec{d}_n^{\,B} &=& \begin{pmatrix} {i\frac{d}{d{\bar N}} \Psi_0} \\{- i\frac{d}{d{\bar N}} \Psi_0} \end{pmatrix} 
\quad{\rm and} \quad
\vec{d}_a^{\,B} = \begin{pmatrix} { \Psi_0} \\{ \Psi_0} \end{pmatrix} 
\eea
where $\Psi_0= \sqrt{\bar{N}} \phi_0$ is the order parameter and $\phi_0$ is the condensate wave function. One has
\bea
{\cal L}_{\rm GP} \,\vec{e}_n^{\,B}  &=& \vec{0}  \\
{\cal L}_{\rm GP} \, \vec{e}_a^{\,B} &=& -i \frac{d\mu}{d\bar{N}} \, \vec{e}_n^{\,B} \,.
\eea
One then obtains the decomposition of unity
\cite{CastinDum1998}
\begin{eqnarray}
\label{eq:decuni}
\openone =  \begin{pmatrix} {\Psi_0} \\{-\Psi_0} \end{pmatrix} 
\begin{pmatrix} {\frac{d}{d{\bar N}} \Psi_0} & {-\frac{d}{d{\bar N}} \Psi_0} \end{pmatrix}
+  \begin{pmatrix} {\frac{d}{d{\bar N}} \Psi_0} \\{\frac{d}{d{\bar N}} \Psi_0} \end{pmatrix}
\begin{pmatrix} {\Psi_0} & {\Psi_0} \end{pmatrix}  \nonumber &&\\
+ \sum_{\lambda} \begin{pmatrix} {U_\lambda^B} \\{V_\lambda^B} \end{pmatrix} 
\begin{pmatrix} {U_\lambda^{B \ast}} & {-V_\lambda^{B \ast}} \end{pmatrix}
+ \begin{pmatrix} {V_\lambda^{B \ast}} \\ {U_\lambda^{B \ast}} \end{pmatrix} 
\begin{pmatrix} {-V_\lambda^B} & {U_\lambda^B} \end{pmatrix} &&
\end{eqnarray}


\paragraph{The adiabatic phase}


The adiabatic phase $\hat{Q}^B$ and all the other operators that have a simple 
linearized dynamics are obtained by projecting the vector of field fluctuations over the modes discussed above, using their dual vectors given in (\ref{eq:decuni}):
\begin{multline}
\begin{pmatrix} { \delta \hat{\psi}} \\ {\delta \hat{\psi}^\dagger} \end{pmatrix} =
\begin{pmatrix} { i \Psi_0} \\ {-i \Psi_0} \end{pmatrix}  \hat{Q}^B
+ \begin{pmatrix} {\frac{d}{d{\bar N}} \Psi_0} \\ {\frac{d}{d{\bar N}} \Psi_0} \end{pmatrix}  \hat{P}^B  \\
\!\!\!\! + \sum_{\lambda} \begin{pmatrix} {U_\lambda^B} \\ {V_\lambda^B} \end{pmatrix} \hat{B}_\lambda
+ \begin{pmatrix} {V_\lambda^{B \ast}} \\ {U_\lambda^{B \ast}} \end{pmatrix} \hat{B}^\dagger_\lambda
\label{eq:fluct_expanded}
\end{multline}
with the operator valued coefficients
\bea
\hat{Q}^B &=& - i \sum_\rr b^3 \frac{d}{d{\bar N}}\Psi_0(\rr) \left[ \delta \hat{\psi}(\rr) - \delta \hat{\psi}^\dagger(\rr) \right]  
\label{eq:QB} \\
\hat{P}^B &=& \sum_\rr b^3 \Psi_0(\rr) \left[ \delta \hat{\psi}(\rr) + \delta \hat{\psi}^\dagger(\rr) \right]  = \delta \hat{N} 
\label{eq:PB} \\
\hat{B}_\lambda &=&  \sum_\rr b^3  \left[ U_\lambda^{B \ast}(\rr)  \, \delta \hat{\psi}(\rr) - V_\lambda^{B \ast}(\rr) \, \delta \hat{\psi}^\dagger(\rr) \right].
\eea
From \eqref{eq:ldB} and \eqref{eq:fluct_expanded} we easily get
\be
\frac{d}{dt}{\hat{Q}}^B = - \frac{1}{\hbar} \frac{d\mu}{d\bar{N}} \hat{P}^B \quad \quad \frac{d}{dt}{\hat{P}}^B = 0 
\ee
As in the fermionic case one can check that $\hat{Q}^B$ is a generator of adiabatic translations of the number of particles
\be
e^{i \delta N \hat{Q}^B} \ket{\alpha(\bar{N})} \underset{\delta N \ll \bar{N}}{\simeq} \ket{\alpha(\bar{N}{-\delta N})}
\label{eq:translation_B}
\ee
where $\ket{\alpha(\bar{N})}$ is the bosonic coherent state
\be
\ket{\alpha(\bar{N})} =  e^{\sum_\rr b^3 [\Psi_0(\bar{N},\rr) \, \hat{\psi}^\dagger(\rr) - \mathrm{h.c.}] } \ket{0}
\ee
{To show \eqref{eq:translation_B}, one can use the identity
\be
e^{\hat{A}} e^{\hat{B}}=e^{\hat{A}+\hat{B}} e^{\frac{1}{2}[\hat{A},\hat{B}]}
\ee 
where $\hat{A}$ and $\hat{B}$ are arbitrary linear combinations of $\hat{\psi}^\dagger(\rr)$ and $\hat{\psi}(\rr)$.
}

\paragraph{Phase of the condensate}


We define the phase of the condensate as the phase of the  operator $\hat{a}_0^B$ that annihilates a particle in the condensate mode $\phi_0$. To first order in the field fluctuations:
\be
\hat{\theta}_0^B = \frac{1}{2i\sqrt{\bar{N}}} \sum_\rr b^3 \phi_0(\rr) \left[ \delta \hat{\psi}(\rr) - \delta \hat{\psi}^\dagger(\rr) \right]
\label{eq:thetaB}
\ee
From the bosonic commutation relation $[\hat{a}_0^B,(\hat{a}_0^B)^\dagger]=1$ it results that $\hat{\theta}_0^B$ is conjugate to $\hat{N}_0$. As a consequence $\hat{\theta}_0^B$ is the generator of translations of the number of 
condensed particles \cite{CastinSinatra2012}. At variance with the transformation induced by $\hat{Q}_0^B$, the transformation induced by $\hat{\theta}_0^B$ is non adiabatic as soon as the condensate 
wavefunction depends on $N$.

Indeed from the definition of the phases (\ref{eq:thetaB}) and (\ref{eq:QB}), and given the fact that
\be
\frac{d}{d{\bar N}} \Psi_0 = \frac{\phi_0}{2 \sqrt{\bar N}} + \sqrt{\bar N} \, \frac{d}{d{\bar N}} \phi_0
\ee
we see 
the adiabatic phase and the phase of the condensate are different operators if $\frac{d}{d{\bar N}} \phi_0 \neq 0$.
In particular we have
\be
\hat{\theta}_0^B = \hat{Q}^B - \frac{1}{i} \sum_{\lambda} \bb{\alpha_\lambda \hat{B}_\lambda - \alpha_\lambda^* \hat{B}_\lambda^\dagger}
\label{eq:QB_theta0}
\ee
with
\bea
\alpha_\lambda&=&  -\frac{1}{2 \sqrt{\bar N}}\, \sum_\rr b^3 \phi_0(\rr)  [U_\lambda^B(\rr) -V_\lambda^B(\rr)] \nonumber \\
&=& \sqrt{\bar N} \, \sum_\rr b^3 \frac{d}{d{\bar N}} \phi_0(\rr) \left[ U_\lambda^B(\rr) -V_\lambda^B(\rr) \right] 
\eea	
where we inserted \eqref{eq:fluct_expanded} in {eq:thetaB} and use the fact that $\sum_\rr b^3 \frac{d}{d{\bar N}} \Psi_0 \bb{U_\lambda^B-V_\lambda^B}=0$.
By taking the time derivative of \eqref{eq:QB_theta0} we obtain the dynamics of the phase of the condensate
$\hat{\theta}_0^B$
\be
\frac{d}{dt}{\hat{\theta}}_0^B = - \frac{1}{\hbar} \frac{d\mu}{dN} \hat{P}^B 
+ \frac{1}{\hbar} \sum_{\lambda} \bb{\alpha_\lambda \epsilon_\lambda^B \hat{B}_\lambda + \mbox{h.c.}}
\label{eq:theta0_dot}
\ee
The operators $\hat{B}_\lambda$ corresponding to excited Bogoliubov modes oscillate in time. The equation 
\eqref{eq:theta0_dot} then indicates that, although they are different operators, ${\hat{\theta}}_0^B$ and $\hat{Q}^B$ have
the same dynamics at long times, dominated by the constant term in \eqref{eq:theta0_dot} proportional to $\hat{P}^B$.
The oscillating terms in the condensate phase derivative \eqref{eq:theta0_dot} are also found within a number conserving theory in the spatially inhomogeneous case \cite{EPL}. 


\section{Beyond BCS theory}


\label{sec:beyond}


\subsection{Restoring a physical state by mixing symmetry breaking states}


\label{subsec:restoringsymmetry}

In our study of the phase dynamics, the choice of a symmetry-breaking ground state as the starting point of a linearized treatment is more than a matter of convenience in the sense that no zero temperature phase dynamics would appear in a state with a fixed number of particle. Therefore we need to give a precise meaning to this choice of an \textit{a priori} non physical state.

Experimentally one cannot prepare a BCS state with a well defined phase. The choice of a broken symmetry state is however meaningful when we deal with a bi-partite system, for example a Josephson junction with two wells $a$ and $b$. We imagine that the two subsystems interact in a way that fixes the relative phase $\phi$ but leaves the global phase $\theta$ as a free parameter.
For a given realization we write the initial state of the system as a product of BCS ground states:
\be
\ket{\psi(\theta)}=\mathcal{N}_a \mathcal{N}_b e^{e^{i\theta}\bb{\alpha \hat{C}_a^\dagger+e^{i\phi}\beta \hat{C}_b^\dagger}}\ket{0}
\label{eq:psi_theta}
\ee
where $\alpha$ and $\beta$ are real numbers and $\hat{C}_{a(b)}^\dagger$ creates a pair in well $a(b)$ centered in
position $\rr_{a(b)}$. As we said, the relative phase $\phi$ is fixed whereas the global phase $\theta$ must be treated as a random variable. The physical state of the system is then described by the density operator
\be
\hat{\rho}=\int_0^{2\pi}  \frac{d\theta}{2\pi} \ket{\psi(\theta)}\bra{\psi(\theta)}
\ee
since, contrarily to (\ref{eq:psi_theta}), it does not contain unphysical coherences between states of different total numbers of particles
\cite{CastinDalibard1997}.
The corresponding  correlation function
\be
C(t)=\Tr{\hat{\rho} \hat{\psi}_\uparrow^\dagger(\rr_b,t) \hat{\psi}_\downarrow^\dagger(\rr_b,t)\hat{\psi}_\downarrow(\rr_a,t){\hat{\psi}_\uparrow}(\rr_a,t)}
\ee
can be expressed in terms of the order parameters of the two subsystems:
\begin{multline}
C(t)=\bra{\psi_{\rm BCS}^b} \hat{\psi}_\uparrow^\dagger(\rr_b,t) \hat{\psi}_\downarrow^\dagger(\rr_b,t) \ket{\psi_{\rm BCS}^b} \\ \times \bra{\psi_{\rm BCS}^a} \hat{\psi}_\downarrow(\rr_a,t) \hat{\psi}_\uparrow(\rr_a,t) \ket{\psi_{\rm BCS}^a}
\end{multline}
where $\ket{\psi_{\rm BCS}^a}=\mathcal{N}_a e^{\alpha \hat{C}_a^\dagger}\ket{0}$ and $\ket{\psi_{\rm BCS}^b}=\mathcal{N}_b e^{e^{i\phi}\beta \hat{C}_b^\dagger}\ket{0}$.
The predicted Gaussian collapse (\ref{eq:Delta_t}) of these order parameters thus implies a Gaussian decay of the correlation function.


\subsection{Blurring time beyond mean field approximation}


\label{subsec:beyondmeanfield}

\subsubsection{Collapse of the order parameter beyond the mean field approximation}
Now that we have given our symmetry breaking description of phase dynamics a precise meaning, we wish to generalize it beyond the mean field approximation. To this end we introduce the state $|\psi\rangle$: 
\begin{equation}
\label{eq:psi}
|\psi\rangle = \sum_N c_N \ket{\psi_0(N)}
\end{equation}
where $\ket{\psi_0(N)}$ is the \textit{exact ground state} with exactly $N$ particles, and $c_N$ are coefficients of a distribution peaked around $\bar{N}$. The state (\ref{eq:psi}) differs from the BCS ground state insofar as the projection of the  BCS state of \eqref{eq:BCSstate} onto the subspace with $N$ particles is not the exact ground state for that particle number. The state $|\psi\rangle$ leads to a non zero order parameter:
\begin{equation} 
\Psi\equiv\bra{\psi} \hat{\psi}_\downarrow \hat{\psi}_\uparrow \ket{\psi} \neq 0
\end{equation} 
that we will show to undergo a collapse using minimal approximations.

During the time evolution, each eigenstate $\ket{\psi_0(N)}$ in \eqref{eq:psi} acquires a phase factor involving its energy $E_0(N)$ in the canonical ensemble, that we expand around $N=\bar{N}$:
\begin{equation}
\label{eq:energiegeneral}
E_0(N)-\mu N=E_0(\bar{N})  -\mu \bar{N} + \frac{1}{2}  (N-\bar{N})^2  \frac{d\mu}{dN}(N=\bar{N}) + \ldots 
\end{equation}
For the time evolution of the order parameter this implies
\begin{multline}
\Psi(t) = e^{2i\frac{t}{\hbar}(-\mu+\frac{d\mu}{dN})} \sum_N \exp\left[-\frac{2it}{\hbar} (N-\bar{N}) \frac{d\mu}{dN}  \right] \\ \times   
\bra{\psi_0(N-2)}\hat{\psi}_\downarrow \hat{\psi}_\uparrow \ket{\psi_0(N)}  c_{N-2}^*c_N
\end{multline}
Assuming a weak dependence of the matrix element on $N$:
\begin{multline}
\label{eq:weakdependance}
\bra{\psi_0(N-2)}\hat{\psi}_\downarrow \hat{\psi}_\uparrow \ket{\psi_0(N)} \simeq \bra{\psi_0(\bar{N}-2)}\hat{\psi}_\downarrow \hat{\psi}_\uparrow \ket{\psi_0(\bar{N})} \\ \simeq \Psi(t=0) 
\end{multline}
and changing the sum into an integral for a Gaussian distribution of $c_N$ of width $\sqrt{{\rm Var}\hat{N}}$, one obtains the expected collapse
\begin{equation}
\Psi(t) =\Psi(0) \exp\left\{-2\, {\rm Var}\hat{N} \left(\frac{d \mu}{dN} \right)_{N=\bar{N}}^2 \frac{t^2}{\hbar}\right\}
\label{eq:Psi_t}
\end{equation}

From (\ref{eq:Psi_t}), we recover the expression of the blurring time $t_{\rm br}$ of \eqref{eq:tbr},
except that now the {\it exact chemical potential} and the {\it physical variance} of the particle number enter.

\subsubsection{Generalization of the adiabatic phase and its dynamics}
\label{seq:facteurrphase}
We now generalize the adiabatic phase and show that its dynamics can be formally derived beyond the linearized dynamics of Sec. \ref{sec:collapse}.  
\paragraph{Generalized adiabatic phase}
We define the generalized adiabatic phase $\hat{Q}$ as a generator of translations of the exact ground states:
\be
\label{eq:phaseadiabatic}
e^{i \delta N \hat{Q}}\ket{\psi_0(N)}=\ket{\psi_0(N{-}2\delta N)}
\ee
Using this definition, we can calculate the commutators:
\begin{multline}
e^{-i \hat{Q}} [\hat{H}, e^{i \hat{Q}} ] = \sum_N [E_0(N-2)-E_0(N)] \ket{\psi_0(N)}\bra{\psi_0(N)} \\ \equiv {-}2 \mu(\hat{N})
\end{multline}
and
\be
e^{-i \hat{Q}} \underbrace{[\hat{H}, [\hat{H},\ldots [\hat{H},}_{p}e^{i \hat{Q}} ] = \bb{{-}2\mu(\hat{N})}^p
\ee
one can show that formally
\be
\label{eq:adiabaticphase}
e^{i \hat{Q}(t)}=e^{i \hat{Q}{(0)}}e^{{-}2i \mu(\hat{N}) \frac{t}{\hbar}}
\ee
A more convenient expression can be obtained at short times, writing $\hat{Q}(t)=\hat{Q}(0) + \hat{Q}(t)-\hat{Q}(0)$ and 
treating $\hat{Q}(t)-\hat{Q}(0)$ as an infinitesimal \footnote{The non-commutation of $\hat{Q}(t)-\hat{Q}(0)$ with $\hat{Q}(0)$ introduces
a fuzziness on $\hat{N}$ in the right-hand side of (\ref{eq:evolQ}) at most of order unity and negligible for large particle numbers.}:
\be
\label{eq:evolQ}
\hat{Q}(t)-\hat{Q}{(0)}\simeq {-}2\mu(\hat{N}) \frac{t}{\hbar}
\ee
whose variance predicts a $t^2$ ballistic spreading of the phase change during $t$, that reproduces
\eqref{eq:tbr} for small relative fluctuations of the total atom number.
Another application of \eqref{eq:adiabaticphase} is that, in the case of negligible initial fluctuations
in the phase operator $\hat{Q}$ around zero,
\be
\meanv{e^{i\hat{Q}(t)}} \simeq \meanv{e^{{-}2i\mu(\hat{N}) \frac{t}{\hbar } } }
\ee
that reproduces {the Gaussian decay of} \eqref{eq:Psi_t}  for weak relative fluctuations of $\hat{N}$.


\section{Hints for an experiment}


\label{sec:hintsforanexperiment}

To conclude this work, we propose an experimental situation in which one could observe, for any value of the interaction strength, the phase dynamics that we have described.
The first step is to prepare an {\it ideal} gas of bosonic dimers in a trap (see Fig.~5).
\begin{figure}[h!]
 \includegraphics[scale=0.25]{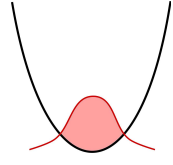}
\caption{\label{fig:puitsimple} Ideal gas of bosonic dimers initially in a single-well trap}
\end{figure}
In the middle of this trap we adiabatically raise a barrier that splits it into two wells $a$ and $b$  (see Fig.~6). 
{Such a splitting was achieved experimentally in \cite{Kohstall2011}}.
\begin{figure}[h!]
 \includegraphics[scale=0.25]{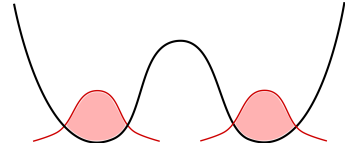}
\caption{\label{fig:puitdouble} The gas is split into two components by adiabatically changing the single-well potential of Fig.~\ref{fig:puitsimple} to a double-well potential}
\end{figure} 
During this stage, the tunneling link and the adiabatic variation of the trapping potential ensure that the gas remains in its ground state:
\begin{eqnarray}
\ket{\psi}&=&\frac{1}{\sqrt{(N/2)!}}\bb{\frac{\hat{a}^\dagger+\hat{b}^\dagger}{\sqrt{2}}}^{N/2} \ket{0} \\
&=&\sum_{n_a=0}^{N/2} c_{n_a} \ket{n_a:\phi_a;N/2-n_a:\phi_b}
\end{eqnarray}
where {$c_{n_a}=\frac{1}{\sqrt{2^{N/2}}} \binom{N/2}{n_a}^{1/2}$}, $\phi_{a(b)}$ is the condensate wave function in the $a(b)$ well,
$n_a$ is the number of bosonic dimers in the $a$ well, and $N$ is the total number of \textit{fermions} in the two wells.

We then cut the link between the two wells so that they form isolated but entangled systems. We tune the scattering length to reach the region of the BEC-BCS crossover we wish to study. The evolution is slow enough so that the state $\ket{n_a:\phi_a;N/2-n_a:\phi_b}$ evolves to a state $\ket{\psi_0^a(2n_a);\psi_0^b(N-2n_a)}$ with 
$2 n_a$ fermions ($N-2 n_a$ fermions) in the ground state of the $a \ (b)$ well:
\begin{eqnarray}
\ket{\psi}=\sum_{n_a=0}^{N/2} c_{n_a} \ket{\psi_0^a(2n_a);\psi_0^b(N-2n_a)}
\label{eq:stateafterFeshbach}
\end{eqnarray}
but fast enough so that phase dynamics during this {stage} can be neglected.
We then let the the state \eqref{eq:stateafterFeshbach} evolve from $0$ to $t$.

Finally, we measure the left-right equal time correlation function
\begin{eqnarray}
C_{ab}(t)&=&\meanv{\hat{\psi}_\uparrow^\dagger(\rr_b,t) \hat{\psi}_\downarrow^\dagger(\rr_b,t) \hat{\psi}_\downarrow(\rr_a,t) \hat{\psi}_\uparrow(\rr_a,t)} \notag
\end{eqnarray}
Following the steps of  the derivation of the previous section (Sec. \ref{seq:facteurrphase}), and writing with a subscript $a(b)$ the operators acting on the $a(b)$ well, we predict
\be
C_{ab}(t)=C_{ab}(0) \exp\left[-\frac{2t^2}{\hbar^2} \bb{\frac{d \mu}{dN}}^2 \text{Var\,}\bb{\hat{N}_a-\hat{N_b}}\right]
\ee
where the second equality assumes the quadratic expansion of the energy around the mean atom numbers {for equal chemical potential in the two wells}.

Direct measurement of pair correlation functions using noise correlations in time-of-flight images have been proposed in \cite{Altman2004,CarusottoCastin2005,Gritsev2008,AspectDemler2011}.  Experimental results in that direction have been obtained for fermions in \cite{GreinerRegal2005,Bloch2006}. An alternative possibility is to ramp the interactions back to the BEC side before the measurement. The pair correlation functions should behave as bosonic one-body correlations functions whose measurement was already achieved.


\section{Conclusion}


\label{sec:conclusion}

By linearization of the equations of motion around the BCS ground state for an interacting spin $1/2$ Fermi gas, both in a {time-dependent Bogoliubov-de Gennes} approach and in the RPA approach, we microscopically derived the existence of and we gave the explicit expressions for a zero-energy mode and an anomalous mode, associated to infinitesimal generators of the translation of the phase and of the adiabatic translation of the number of particles, respectively. Projection of the quantum field of pairs on these modes yields conjugate phase and number operators, and the linearly increasing dispersion of the phase operator in time is responsible for the collapse of the order parameter. We predict a coherence time of the order parameter depending of the derivative of the chemical potential of the gas with respect to the atom number, and on the variance of that atom number. In the thermodynamic limit, with a variance scaling as the mean atom number,
the coherence time scales as the square root of the system size, and it is thus observable in systems with relatively small particle numbers, such as cold atomic gases. 
As expected, in the BEC limit of the crossover, where the ground state of the system is a condensate of bosonic dimers, 
our formula for the coherence time is consistent with what is known for bosons.
 
Further studying our phase operator, we interpret it as a generator of adiabatic particle number translations for the ground state of the gas; it is in general different from the phase operator of the condensate defined from the amplitude of the field of pairs on the condensate mode. This difference originates from a dependence of the condensate wave function on the number of particles, which is the case for our fermionic system, even in the spatially homogeneous case. With this interpretation of the phase operator in mind, we were able to extend our results for the blurring time and for the phase dynamics beyond the BCS mean field approximation.


\section*{Acknowledgments}


We thank C\'eline Nadal for her contribution \cite{Nadal} at an early stage of the project and for useful discussions.
We acknowledge financial support from the QIBEC European project.


\appendix


\section{Relation between the {linearized time-dependent Bogoliubov-de Gennes} and the RPA equations}


\label{app:equations_semi}


\subsection{{Time-dependent Bogoliubov-de Gennes} equations}


From the expression \eqref{eq:Eclass}  {it is apparent that the derivative of the energy with respect to $V_\kk$
can be deduced from the derivatives of the density and order parameter}
\begin{eqnarray}
\frac{\partial \rho_\uparrow}{\partial V_\kk} &=& \frac{V_\kk^*}{L^3} \\
\frac{\partial \Delta}{\partial V_\kk} &=& -\frac{g_0}{L^3} \bb{U_\kk - \frac{|V_\kk|^2}{2U_\kk}} \\
\frac{\partial \Delta^*}{\partial V_\kk} &=& \frac{g_0}{L^3} \frac{\bb{V_\kk^*}^2}{2U_\kk}
\end{eqnarray} 
where we used \eqref{eq:Uk}, \eqref{eq:defDelta} and \eqref{eq:rhoup}.
We then obtain
\begin{multline}
{-i\hbar \frac{d}{dt} V_\kk^*} = \frac{\partial E}{\partial V_\kk} = 2\bb{\frac{\hbar^2 k^2}{2m} -\mu + g_0 \rho_\uparrow }V_\kk^*  \\ -
\Delta^* U_\kk + \frac{V_\kk^*}{2U_\kk} \bb{V_\kk \Delta^* + V_\kk^* \Delta }
\end{multline}
By linearizing around the BCS ground state value $V_\kk^0$, we obtain
\begin{multline}
- i\hbar \frac{d}{dt} \delta V_\kk^* =  \frac{\partial E}{\partial V_\kk} \simeq\bb{ 2 \xi_\kk + \frac{\Delta_0 V_\kk^0}{U_\kk^0}   } \delta V_\kk^* \\ +
 \frac{\Delta_0 V_\kk^0}{U_\kk^0} \bb{1+ \frac{(V_\kk^0)^2}{2(U_\kk^0)^2}}\bb{ \delta V_\kk+ \delta V_\kk^*} \\ +
 2 g_0 V_\kk^0 \delta \rho_\uparrow - U_\kk^0 \delta \Delta^* + \frac{(V_\kk^0)^2}{2U_\kk^0} 
 \bb{\delta \Delta + \delta \Delta^*} \label{eq:Vkstarpoint}
\end{multline}
where $\delta \rho_\uparrow$ and $\delta \Delta$ are obtained by linearizing \eqref{eq:defDelta} and \eqref{eq:rhoup}
\bea
\!\!\!\!\! \!\! \delta \rho_\uparrow \!\! &=&	\!\!  \frac{1}{L^3} \sum_\qq V_\qq^0 \bb{ \delta V_\qq+ \delta V_\qq^*} \\
\!\!\!\!\! \!\! \delta \Delta \!\!  &=& \!\!  -\frac{g_0}{L^3} \sum_\qq \left[ \bb{U_\qq^0 - \frac{(V_\qq^0)^2}{2U_\qq^0} }\delta V_\qq
 - \frac{(V_\qq^0)^2}{2U_\qq^0}  \delta V_\qq^* \right]
\eea
Referring to the notations in \eqref{eq:Lblock}, we finally obtain
\begin{multline}
A_{\kk \qq} = \delta_{\kk \qq} \left\{ 2 \xi_\kk +  \frac{\Delta_0 V_\kk^0}{U_\kk^0} \left[2+ \frac{(V_\kk^0)^2}{2(U_\kk^0)^2}\right] \right\} \\
+ {\frac{g_0}{L^3} \left[ 2 \bb{U_\kk^0 U_\qq^0 + V_\kk^0 V_\qq^0}  + 
\frac{(V_\kk^0 V_\qq^0)^2 - (U_\kk^0)^2 - (U_\qq^0)^2}{2 U_\kk^0 U_\qq^0}\right] }
\end{multline}
\begin{multline}
B_{\kk \qq} = \delta_{\kk \qq}   \frac{\Delta_0 V_\kk^0}{U_\kk^0} \left[1+ \frac{(V_\kk^0)^2}{2(U_\kk^0)^2}\right] \\
+ {\frac{g_0}{L^3} \left[ U_\kk^0 U_\qq^0 + 2 V_\kk^0 V_\qq^0 + 
\frac{(V_\kk^0 V_\qq^0)^2 - (U_\kk^0)^2 - (U_\qq^0)^2}{2 U_\kk^0 U_\qq^0}
 \right] }
\end{multline}



\subsection{Link with the RPA equations}


Using the correspondence \eqref{eq:corr_y} and  \eqref{eq:corr_m} 
{and \eqref{eq:Vkstarpoint}} we obtain
\bea
 i\hbar \frac{d}{dt} \hat{m}_\kk  &\leftrightarrow&  
2 \Delta_0 U_\kk^0 \bb{\delta V_\kk - \delta V_\kk^*} - 2U_\kk^0 V_\kk^0 \bb{\delta \Delta - \delta \Delta^*} \nonumber \\
&& \label{eq:mkp_sc} \\
i\hbar \frac{d}{dt} \hat{y}_\kk & \leftrightarrow &
- \frac{4\epsilon_\kk^2 V_\kk^0}{\Delta_0} \bb{\delta V_\kk + \delta V_\kk^*} - 4 g_0 V_\kk^0 U_\kk^0 \delta \rho_\uparrow
\nonumber \\&+& \left[(U_\kk^0)^2-(V_\kk^0)^2\right] \bb{\delta \Delta + \delta \Delta^*}
\label{eq:ykp_sc}
\eea
where, {to obtain the simplified forms of the coefficients} of $\delta V_\kk - \delta V_\kk^*$ in \eqref{eq:mkp_sc} and 
of $\delta V_\kk + \delta V_\kk^*$ in \eqref{eq:ykp_sc},
we used the equation \eqref{eq:gap} in the forms
\be
\Delta_0 V_\kk^0 = \bb{\epsilon_\kk - \xi_\kk} U_\kk^0 \qquad \Delta_0 U_\kk^0 = \bb{\epsilon_\kk + \xi_\kk} V_\kk^0
\ee
{or alternatively we eliminated $\xi_\kk$ from \eqref{eq:Vkstarpoint} using \eqref{eq:utile3}}.
Remarking that
\bea
\delta \Delta + \delta \Delta^* &\leftrightarrow& - \frac{g_0}{L^3} \sum_\kk \frac{\xi_\kk}{\Delta_0} \hat{m}_\kk \\
\delta \Delta - \delta \Delta^* &\leftrightarrow& \frac{g_0}{L^3} \sum_\kk \hat{y}_\kk \\
\delta \rho_\uparrow  &\leftrightarrow& \frac{1}{2L^3} \sum_\kk \hat{m}_\kk
\eea
{and} using \eqref{eq:meanb} we then {recover the equations of motion}
\eqref{eq:RPAexpli1} and \eqref{eq:RPAexpli2} except for the $\hat{S}_\kk$ contribution.
{This is not surprising because the BCS ground state is an eigenstate of $\hat{S}_\kk$ with zero eigenvalue
(see \eqref{eq:zetak_psi0}).
The expectation value $\langle \hat{S}_\kk \rangle_{\rm BCS}$ is then a quantity of second order 
in $\delta V_\kk$, $\delta V_\kk^*$}.


\section{Computing $\meanv{e^{i\hat{Q}{(t)}}}$ in the BCS approximation}


\label{sec:appendixcalculcollapse}

In section \ref{sec:collapse} we obtained
\be
\hat{Q}(t)=\sum_\kk \hat{Q}_\kk(t)
\label{eq:appendix1}
\ee
where $\hat{Q}_\kk(t)$ is a linear combination of the operators $\hat{m}_\kk(0)$, $\hat{y}_\kk(0)$ and $\hat{\zeta}_\kk$
{with time-dependent coefficients.} Due to the factorized form of the BCS state, we have
\be
\meanv{e^{i\hat{Q}(t)}}_0
=\prod_\kk\bra{\psi_\kk}e^{i\hat{Q}_\kk(t)}\ket{\psi_\kk} \label{eq:appendix2}
\ee
where $\ket{\psi_\kk}=\bb{U_\kk^0-V_\kk^0\hat{d}_\kk^\dagger}\ket{0}$. 
Within the two-dimensional subspace spanned by $\ket{0_\kk}=\ket{0}$ and 
$\ket{1}_\kk=\hat{d}_\kk^\dagger\ket{0}$, the operator $\hat{Q}_\kk$ acts as a linear superposition 
of Pauli matrices and $\openone$, which allows to calculate its exponential exactly \cite{Nadal}.
If we however consider (i) the thermodynamic limit and (ii) time scales of order $L^{3/2}$ (as expected
from the expression of $t_{\rm br}$) such that $t/L^3\to 0$ for $L\to \infty$, we find that $\hat{Q}_\kk(t) \to 0$
and it is sufficient to expand its exponential to second order {\it within each factor} of the product over $\kk$:
\be
\meanv{e^{i\hat{Q}(t)}}_0 \simeq\prod_\kk \left[ 1 - \frac{1}{2} \meanv{\hat{Q}_\kk^2(t)} \right]
\simeq e^{- \meanv{\hat{Q}^2(t)}/2} \,
\ee
where we use the fact that $\sum_\kk \meanv{\hat{Q}_\kk^2(t)}=\meanv{\hat{Q}^2(t)}$.
The contribution of $\hat{Z}$ to the variance of $\hat{Q}(t)$ is exactly zero due to \eqref{eq:Z0} as well
as the one of the crossed terms between $\hat{Q}(0)$ and $\hat{P}$, so that
\be
\meanv{\hat{Q}^2(t)} = \meanv{\hat{Q}^2(0)} + \frac{4t^2}{\hbar^2} \bb{\frac{d\mu}{d{\bar N}}}^2 \meanv{\hat{P}^2}
\ee
By legitimately neglecting the variance of $\hat{Q}(0)$, 
\be
\text{Var\,} \hat{Q}{(0)}= 4 \sum_\kk \frac{\bb{\frac{dV_\kk^0}{d {\bar N}}}^2}{(U_\kk^{0})^2} = O\bb{\frac{1}{N}}
\label{eq:varQ0}
\ee
we finally recover \eqref{eq:Delta_t}.


{\section{{Initial} variance of the adiabatic phase in the BCS ground state}


\label{app:varQ}

{In equation \eqref{eq:varQ0}},
$\frac{dV_\kk^0}{d \mu}$ can be deduced from the first line of \eqref{eq:interm2},
\be
\frac{dV_\kk^0}{d \mu}=\frac{\Delta_0 U_\kk^0}{2\epsilon_\kk^2}\bb{1+\frac{X}{\Theta}\frac{\xi_\kk}{\Delta_0}}
\ee
where we have neglected the mean field contribution $g_0 d\rho_\uparrow^0/d\mu$ {in the $b\to0$ limit} and used the relation \eqref{eq:dDelta_dmu}.
Expressing $d\mu/d\bar{N}$ in terms of $X$ and $\Theta$ from \eqref{eq:dmudN_XandTheta}, we obtain
\begin{multline}
\text{Var} \, \hat{Q}(0)=\frac{1}{(X^2+\Theta^2)^2} \bigg[(\Theta^2-X^2) \sum_\kk \frac{\Delta_0^4}{\epsilon_\kk^4} + X^2 \sum_\kk \frac{\Delta_0^4}{\epsilon_\kk^4}\\+2X\Theta\sum_\kk \frac{\Delta_0^3\xi_\kk}{\epsilon_\kk^4}\bigg]
\end{multline}


\section{Details on the computation of $\frac{d}{dt}\hat{\theta_0}$}


\label{app:theta_0}

Taking the temporal derivative of \eqref{eq:phaseofthecond} and replacing $\frac{d}{dt}\hat{y}_\kk$ by its RPA expression leads to
\begin{multline}
\frac{d}{dt}\hat{\theta}_0=-\frac{1}{2\hbar\bar{N}_0}\bigg[ \sum_\kk \bb{\epsilon_\kk +\frac{g_0}{2L^3}\Sigma_1+\frac{g_0 \xi_\kk}{2L^3}\Sigma_2}\hat{m}_\kk \\
-\sum_\kk \frac{\Delta_0}{2\epsilon_\kk}\hat{S}_\kk \bigg] 
\label{eq:dtheta0_appendice}
\end{multline}
where
\bea
\Sigma_1&=&\sum_\kk \frac{\Delta_0^2}{\epsilon_\kk^2}=\text{Var\,} \hat{N} \\
\Sigma_2&=&\sum_\kk \frac{\xi_\kk}{\epsilon_\kk^2}
\eea
$\Sigma_1$ tends to a finite value in the limit of zero lattice spacing, regardless of the system size. Consequently, its contribution in \eqref{eq:dtheta0_appendice} tends to zero:
\be
\frac{g_0}{2L^3} \Sigma_1 \underset{b\rightarrow0} {\rightarrow} 0.
\ee
$\Sigma_2$ does not converge in the limit of zero lattice spacing, but using the gap equation \eqref{eq:BCS2} we can rewrite it as
\be
\Sigma_2= -\frac{2L^3}{g_0} + \Sigma_3
\ee
where $\Sigma_3=\sum_\kk \frac{\xi_\kk-\epsilon_\kk}{\epsilon_\kk^2}$ converges since $|\Sigma_3|\leq\frac{\bar{N}}{\Delta_0}$. As a consequence
\be
\frac{g_0}{2L^3} \Sigma_2 \underset{b\rightarrow0} {\rightarrow} -1
\ee
and hence the value of the limit of $d\hat{\theta}_0/dt$ given in \eqref{eq:dtheta}.
}


\bibliography{paper8}

\providecommand*\hyphen{-}
\begin{thebibliography}{61}
\expandafter\ifx\csname natexlab\endcsname\relax\def\natexlab#1{#1}\fi
\expandafter\ifx\csname bibnamefont\endcsname\relax
  \def\bibnamefont#1{#1}\fi
\expandafter\ifx\csname bibfnamefont\endcsname\relax
  \def\bibfnamefont#1{#1}\fi
\expandafter\ifx\csname citenamefont\endcsname\relax
  \def\citenamefont#1{#1}\fi
\expandafter\ifx\csname url\endcsname\relax
  \def\url#1{\texttt{#1}}\fi
\expandafter\ifx\csname urlprefix\endcsname\relax\def\urlprefix{URL }\fi
\providecommand{\bibinfo}[2]{#2}
\providecommand{\eprint}[2][]{\url{#2}}

\bibitem[{\citenamefont{Hall et~al.}(1998)\citenamefont{Hall, Matthews, Wieman,
  and Cornell}}]{Cornell1998}
\bibinfo{author}{\bibfnamefont{D.}~\bibnamefont{Hall}},
  \bibinfo{author}{\bibfnamefont{M.}~\bibnamefont{Matthews}},
  \bibinfo{author}{\bibfnamefont{C.}~\bibnamefont{Wieman}}, \bibnamefont{and}
  \bibinfo{author}{\bibfnamefont{E.}~\bibnamefont{Cornell}},
  \bibinfo{journal}{Physical Review Letters} \textbf{\bibinfo{volume}{81}},
  \bibinfo{pages}{1543} (\bibinfo{year}{1998}).

\bibitem[{\citenamefont{Orzel et~al.}(2001)\citenamefont{Orzel, Tuchman,
  Fenselau, Yasuda, and Kasevich}}]{Kasevich2001}
\bibinfo{author}{\bibfnamefont{C.}~\bibnamefont{Orzel}},
  \bibinfo{author}{\bibfnamefont{A.}~\bibnamefont{Tuchman}},
  \bibinfo{author}{\bibfnamefont{M.}~\bibnamefont{Fenselau}},
  \bibinfo{author}{\bibfnamefont{M.}~\bibnamefont{Yasuda}}, \bibnamefont{and}
  \bibinfo{author}{\bibfnamefont{M.}~\bibnamefont{Kasevich}},
  \bibinfo{journal}{Science} \textbf{\bibinfo{volume}{291}},
  \bibinfo{pages}{2386} (\bibinfo{year}{2001}).

\bibitem[{\citenamefont{Greiner et~al.}(2002)\citenamefont{Greiner, Mandel,
  H\"ansch, and Bloch}}]{Bloch2002}
\bibinfo{author}{\bibfnamefont{M.}~\bibnamefont{Greiner}},
  \bibinfo{author}{\bibfnamefont{O.}~\bibnamefont{Mandel}},
  \bibinfo{author}{\bibfnamefont{T.}~\bibnamefont{H\"ansch}}, \bibnamefont{and}
  \bibinfo{author}{\bibfnamefont{I.}~\bibnamefont{Bloch}},
  \bibinfo{journal}{Nature (London)} \textbf{\bibinfo{volume}{419}},
  \bibinfo{pages}{51} (\bibinfo{year}{2002}).

\bibitem[{\citenamefont{Dunningham et~al.}(2005)\citenamefont{Dunningham,
  Burnett, and Phillips}}]{Dunningham2005}
\bibinfo{author}{\bibfnamefont{J.}~\bibnamefont{Dunningham}},
  \bibinfo{author}{\bibfnamefont{K.}~\bibnamefont{Burnett}}, \bibnamefont{and}
  \bibinfo{author}{\bibfnamefont{W.}~\bibnamefont{Phillips}},
  \bibinfo{journal}{Phil. Trans. R. Soc.} \textbf{\bibinfo{volume}{363}},
  \bibinfo{pages}{2165} (\bibinfo{year}{2005}).

\bibitem[{\citenamefont{Shin et~al.}(2004)\citenamefont{Shin, Saba, Pasquini,
  Ketterle, Pritchard, and Leanhardt}}]{Shin2004}
\bibinfo{author}{\bibfnamefont{Y.}~\bibnamefont{Shin}},
  \bibinfo{author}{\bibfnamefont{M.}~\bibnamefont{Saba}},
  \bibinfo{author}{\bibfnamefont{T.}~\bibnamefont{Pasquini}},
  \bibinfo{author}{\bibfnamefont{W.}~\bibnamefont{Ketterle}},
  \bibinfo{author}{\bibfnamefont{D.}~\bibnamefont{Pritchard}},
  \bibnamefont{and}
  \bibinfo{author}{\bibfnamefont{A.}~\bibnamefont{Leanhardt}},
  \bibinfo{journal}{Physical Review Letters} \textbf{\bibinfo{volume}{92}},
  \bibinfo{pages}{050405} (\bibinfo{year}{2004}).

\bibitem[{\citenamefont{Jo et~al.}(2007)\citenamefont{Jo, Shin, Will, Pasquini,
  Saba, Ketterle, Pritchard, Vengalattore, and Prentiss}}]{Ketterle2007}
\bibinfo{author}{\bibfnamefont{G.-B.} \bibnamefont{Jo}},
  \bibinfo{author}{\bibfnamefont{Y.}~\bibnamefont{Shin}},
  \bibinfo{author}{\bibfnamefont{S.}~\bibnamefont{Will}},
  \bibinfo{author}{\bibfnamefont{T.~A.} \bibnamefont{Pasquini}},
  \bibinfo{author}{\bibfnamefont{M.}~\bibnamefont{Saba}},
  \bibinfo{author}{\bibfnamefont{W.}~\bibnamefont{Ketterle}},
  \bibinfo{author}{\bibfnamefont{D.~E.} \bibnamefont{Pritchard}},
  \bibinfo{author}{\bibfnamefont{M.}~\bibnamefont{Vengalattore}},
  \bibnamefont{and} \bibinfo{author}{\bibfnamefont{M.}~\bibnamefont{Prentiss}},
  \bibinfo{journal}{Phys. Rev. Lett.} \textbf{\bibinfo{volume}{98}},
  \bibinfo{pages}{030407} (\bibinfo{year}{2007}).

\bibitem[{\citenamefont{Sols}(1994)}]{Sols1994}
\bibinfo{author}{\bibfnamefont{F.}~\bibnamefont{Sols}},
  \bibinfo{journal}{Physica B} \textbf{\bibinfo{volume}{194}},
  \bibinfo{pages}{1389} (\bibinfo{year}{1994}).

\bibitem[{\citenamefont{Wright et~al.}(1996)\citenamefont{Wright, Walls, and
  Garrison}}]{Walls1996}
\bibinfo{author}{\bibfnamefont{E.~M.} \bibnamefont{Wright}},
  \bibinfo{author}{\bibfnamefont{D.~F.} \bibnamefont{Walls}}, \bibnamefont{and}
  \bibinfo{author}{\bibfnamefont{J.~C.} \bibnamefont{Garrison}},
  \bibinfo{journal}{Physical Review Letters} \textbf{\bibinfo{volume}{77}},
  \bibinfo{pages}{2158} (\bibinfo{year}{1996}).

\bibitem[{\citenamefont{Lewenstein and You}(1996)}]{Lewenstein1996}
\bibinfo{author}{\bibfnamefont{M.}~\bibnamefont{Lewenstein}} \bibnamefont{and}
  \bibinfo{author}{\bibfnamefont{L.}~\bibnamefont{You}},
  \bibinfo{journal}{Physical Review Letters} \textbf{\bibinfo{volume}{77}},
  \bibinfo{pages}{3489} (\bibinfo{year}{1996}).

\bibitem[{\citenamefont{Castin and Dalibard}(1997)}]{CastinDalibard1997}
\bibinfo{author}{\bibfnamefont{Y.}~\bibnamefont{Castin}} \bibnamefont{and}
  \bibinfo{author}{\bibfnamefont{J.}~\bibnamefont{Dalibard}},
  \bibinfo{journal}{Physical Review A} \textbf{\bibinfo{volume}{55}},
  \bibinfo{pages}{4330} (\bibinfo{year}{1997}).

\bibitem[{\citenamefont{Villain et~al.}(1997)\citenamefont{Villain, Lewenstein,
  Dum, Castin, You, Imamoglu, and Kennedy}}]{Villain1997}
\bibinfo{author}{\bibfnamefont{P.}~\bibnamefont{Villain}},
  \bibinfo{author}{\bibfnamefont{M.}~\bibnamefont{Lewenstein}},
  \bibinfo{author}{\bibfnamefont{R.}~\bibnamefont{Dum}},
  \bibinfo{author}{\bibfnamefont{Y.}~\bibnamefont{Castin}},
  \bibinfo{author}{\bibfnamefont{L.}~\bibnamefont{You}},
  \bibinfo{author}{\bibfnamefont{A.}~\bibnamefont{Imamoglu}}, \bibnamefont{and}
  \bibinfo{author}{\bibfnamefont{T.}~\bibnamefont{Kennedy}},
  \bibinfo{journal}{Journal of Modern Optics} \textbf{\bibinfo{volume}{44}},
  \bibinfo{pages}{1775} (\bibinfo{year}{1997}).

\bibitem[{\citenamefont{Berrada et~al.}(2013)\citenamefont{Berrada, van Frank,
  B\"{u}cker, Schumm, Schaff, and Schmiedmayer}}]{Berrada2013}
\bibinfo{author}{\bibfnamefont{T.}~\bibnamefont{Berrada}},
  \bibinfo{author}{\bibfnamefont{S.}~\bibnamefont{van Frank}},
  \bibinfo{author}{\bibfnamefont{R.}~\bibnamefont{B\"{u}cker}},
  \bibinfo{author}{\bibfnamefont{T.}~\bibnamefont{Schumm}},
  \bibinfo{author}{\bibfnamefont{J.-F.} \bibnamefont{Schaff}},
  \bibnamefont{and}
  \bibinfo{author}{\bibfnamefont{J.}~\bibnamefont{Schmiedmayer}},
  \bibinfo{journal}{Nature Commun} \textbf{\bibinfo{volume}{4}},
  \bibinfo{pages}{2077} (\bibinfo{year}{2013}).

\bibitem[{\citenamefont{Will et~al.}(2010)\citenamefont{Will, Best, Schneider,
  Hackerm\"{u}ller, L\"{u}hmann, and Bloch}}]{Will2010}
\bibinfo{author}{\bibfnamefont{S.}~\bibnamefont{Will}},
  \bibinfo{author}{\bibfnamefont{T.}~\bibnamefont{Best}},
  \bibinfo{author}{\bibfnamefont{U.}~\bibnamefont{Schneider}},
  \bibinfo{author}{\bibfnamefont{L.}~\bibnamefont{Hackerm\"{u}ller}},
  \bibinfo{author}{\bibfnamefont{D.-S.} \bibnamefont{L\"{u}hmann}},
  \bibnamefont{and} \bibinfo{author}{\bibfnamefont{I.}~\bibnamefont{Bloch}},
  \bibinfo{journal}{Nature} \textbf{\bibinfo{volume}{465}},
  \bibinfo{pages}{197} (\bibinfo{year}{2010}).

\bibitem[{\citenamefont{Gross et~al.}(2008)\citenamefont{Gross, Zibold,
  Nicklas, Est\`eve, and Oberthaler}}]{Oberthaler2010}
\bibinfo{author}{\bibfnamefont{C.}~\bibnamefont{Gross}},
  \bibinfo{author}{\bibfnamefont{T.}~\bibnamefont{Zibold}},
  \bibinfo{author}{\bibfnamefont{E.}~\bibnamefont{Nicklas}},
  \bibinfo{author}{\bibfnamefont{J.}~\bibnamefont{Est\`eve}}, \bibnamefont{and}
  \bibinfo{author}{\bibfnamefont{M.~K.} \bibnamefont{Oberthaler}},
  \bibinfo{journal}{Nature} \textbf{\bibinfo{volume}{464}},
  \bibinfo{pages}{1165} (\bibinfo{year}{2008}).

\bibitem[{\citenamefont{Riedel et~al.}(2010)\citenamefont{Riedel, B\"ohi, Li,
  H\"ansch, Sinatra, and Treutlein}}]{Treutlein2010}
\bibinfo{author}{\bibfnamefont{M.~F.} \bibnamefont{Riedel}},
  \bibinfo{author}{\bibfnamefont{P.}~\bibnamefont{B\"ohi}},
  \bibinfo{author}{\bibfnamefont{Y.}~\bibnamefont{Li}},
  \bibinfo{author}{\bibfnamefont{T.~W.} \bibnamefont{H\"ansch}},
  \bibinfo{author}{\bibfnamefont{A.}~\bibnamefont{Sinatra}}, \bibnamefont{and}
  \bibinfo{author}{\bibfnamefont{P.}~\bibnamefont{Treutlein}},
  \bibinfo{journal}{Nature} \textbf{\bibinfo{volume}{464}},
  \bibinfo{pages}{1170} (\bibinfo{year}{2010}).

\bibitem[{\citenamefont{Kitagawa and Ueda}(1993)}]{Ueda1993}
\bibinfo{author}{\bibfnamefont{M.}~\bibnamefont{Kitagawa}} \bibnamefont{and}
  \bibinfo{author}{\bibfnamefont{M.}~\bibnamefont{Ueda}},
  \bibinfo{journal}{Phys. Rev. A} \textbf{\bibinfo{volume}{47}},
  \bibinfo{pages}{5138} (\bibinfo{year}{1993}).

\bibitem[{\citenamefont{Wineland et~al.}(1994)\citenamefont{Wineland,
  Bollinger, Itano, and Heinzen}}]{Wineland1994}
\bibinfo{author}{\bibfnamefont{D.~J.} \bibnamefont{Wineland}},
  \bibinfo{author}{\bibfnamefont{J.~J.} \bibnamefont{Bollinger}},
  \bibinfo{author}{\bibfnamefont{W.~M.} \bibnamefont{Itano}}, \bibnamefont{and}
  \bibinfo{author}{\bibfnamefont{D.~J.} \bibnamefont{Heinzen}},
  \bibinfo{journal}{Phys. Rev. A} \textbf{\bibinfo{volume}{50}},
  \bibinfo{pages}{67} (\bibinfo{year}{1994}).

\bibitem[{\citenamefont{Sorensen et~al.}(2001)\citenamefont{Sorensen, Duan,
  Cirac, and Zoller}}]{Sorensen:2001}
\bibinfo{author}{\bibfnamefont{A.}~\bibnamefont{Sorensen}},
  \bibinfo{author}{\bibfnamefont{L.}~\bibnamefont{Duan}},
  \bibinfo{author}{\bibfnamefont{J.}~\bibnamefont{Cirac}}, \bibnamefont{and}
  \bibinfo{author}{\bibfnamefont{P.}~\bibnamefont{Zoller}},
  \bibinfo{journal}{Nature} \textbf{\bibinfo{volume}{409}}, \bibinfo{pages}{63}
  (\bibinfo{year}{2001}).

\bibitem[{\citenamefont{Sinatra
  et~al.}(2012{\natexlab{a}})\citenamefont{Sinatra, Dornstetter, and
  Castin}}]{Frontiers:2012}
\bibinfo{author}{\bibfnamefont{A.}~\bibnamefont{Sinatra}},
  \bibinfo{author}{\bibfnamefont{J.-C.} \bibnamefont{Dornstetter}},
  \bibnamefont{and} \bibinfo{author}{\bibfnamefont{Y.}~\bibnamefont{Castin}},
  \bibinfo{journal}{Front. Phys.} \textbf{\bibinfo{volume}{7}},
  \bibinfo{pages}{86} (\bibinfo{year}{2012}{\natexlab{a}}).

\bibitem[{\citenamefont{Sinatra
  et~al.}(2012{\natexlab{b}})\citenamefont{Sinatra, Witkowska, and
  Castin}}]{Sinatra2012}
\bibinfo{author}{\bibfnamefont{A.}~\bibnamefont{Sinatra}},
  \bibinfo{author}{\bibfnamefont{E.}~\bibnamefont{Witkowska}},
  \bibnamefont{and} \bibinfo{author}{\bibfnamefont{Y.}~\bibnamefont{Castin}},
  \bibinfo{journal}{Eur. Phys. J. Special Topics}
  \textbf{\bibinfo{volume}{102}}, \bibinfo{pages}{87}
  (\bibinfo{year}{2012}{\natexlab{b}}).

\bibitem[{\citenamefont{Inguscio et~al.}(2007)\citenamefont{Inguscio,
  W.Ketterle, and Salomon}}]{Varenna_livre}
\bibinfo{editor}{\bibfnamefont{M.}~\bibnamefont{Inguscio}},
  \bibinfo{editor}{\bibnamefont{W.Ketterle}}, \bibnamefont{and}
  \bibinfo{editor}{\bibfnamefont{C.}~\bibnamefont{Salomon}}, eds.,
  \emph{\bibinfo{title}{Ultra-cold Fermi Gases}}
  (\bibinfo{publisher}{Societ{\`a} italiana di fisica},
  \bibinfo{address}{Bologna, Italia}, \bibinfo{year}{2007}).

\bibitem[{\citenamefont{Giorgini et~al.}(2008)\citenamefont{Giorgini,
  Pitaevskii, and Stringari}}]{Stringari2008}
\bibinfo{author}{\bibfnamefont{S.}~\bibnamefont{Giorgini}},
  \bibinfo{author}{\bibfnamefont{L.~P.} \bibnamefont{Pitaevskii}},
  \bibnamefont{and}
  \bibinfo{author}{\bibfnamefont{S.}~\bibnamefont{Stringari}},
  \bibinfo{journal}{Rev. Mod. Phys.} \textbf{\bibinfo{volume}{80}},
  \bibinfo{pages}{1215} (\bibinfo{year}{2008}).

\bibitem[{\citenamefont{Zwerger}(2012)}]{Zwerger2012}
\bibinfo{editor}{\bibfnamefont{W.}~\bibnamefont{Zwerger}}, ed.,
  \emph{\bibinfo{title}{{The BCS-BEC Crossover and the Unitary Fermi Gas}}}
  (\bibinfo{publisher}{Springer Verlag}, \bibinfo{year}{2012}).

\bibitem[{\citenamefont{Castin}(2007)}]{Varenna}
\bibinfo{author}{\bibfnamefont{Y.}~\bibnamefont{Castin}}, in
  \emph{\bibinfo{booktitle}{Ultra-cold Fermi Gases}}, edited by
  \bibinfo{editor}{\bibfnamefont{M.}~\bibnamefont{Inguscio}},
  \bibinfo{editor}{\bibnamefont{W.Ketterle}}, \bibnamefont{and}
  \bibinfo{editor}{\bibfnamefont{C.}~\bibnamefont{Salomon}}
  (\bibinfo{publisher}{Societ{\`a} italiana di fisica},
  \bibinfo{address}{Bologna, Italia}, \bibinfo{year}{2007}).

\bibitem[{\citenamefont{Randeria and Taylor}(2014)}]{Randeria2014}
\bibinfo{author}{\bibfnamefont{M.}~\bibnamefont{Randeria}} \bibnamefont{and}
  \bibinfo{author}{\bibfnamefont{E.}~\bibnamefont{Taylor}},
  \bibinfo{journal}{Ann. Rev. Cond. Mat. Phys.} \textbf{\bibinfo{volume}{5}},
  \bibinfo{pages}{065027} (\bibinfo{year}{2014}).

\bibitem[{\citenamefont{Zwierlein et~al.}(2005)\citenamefont{Zwierlein,
  Abo-Shaeer, Schirotzek, Schunck, and Ketterle}}]{Ketterle2005}
\bibinfo{author}{\bibfnamefont{M.~W.} \bibnamefont{Zwierlein}},
  \bibinfo{author}{\bibfnamefont{J.~R.} \bibnamefont{Abo-Shaeer}},
  \bibinfo{author}{\bibfnamefont{A.}~\bibnamefont{Schirotzek}},
  \bibinfo{author}{\bibfnamefont{C.~H.} \bibnamefont{Schunck}},
  \bibnamefont{and} \bibinfo{author}{\bibfnamefont{W.}~\bibnamefont{Ketterle}},
  \bibinfo{journal}{Nature} \textbf{\bibinfo{volume}{435}},
  \bibinfo{pages}{1047} (\bibinfo{year}{2005}).

\bibitem[{\citenamefont{Sidorenkov et~al.}(2013)\citenamefont{Sidorenkov, Tey,
  Grimm, Hou, and Pitaevskii}}]{Grimm:2013}
\bibinfo{author}{\bibfnamefont{L.~A.} \bibnamefont{Sidorenkov}},
  \bibinfo{author}{\bibfnamefont{M.~K.} \bibnamefont{Tey}},
  \bibinfo{author}{\bibfnamefont{R.}~\bibnamefont{Grimm}},
  \bibinfo{author}{\bibfnamefont{Y.-H.} \bibnamefont{Hou}}, \bibnamefont{and}
  \bibinfo{author}{\bibfnamefont{L.}~\bibnamefont{Pitaevskii}},
  \bibinfo{journal}{Nature} \textbf{\bibinfo{volume}{498}}, \bibinfo{pages}{78}
  (\bibinfo{year}{2013}).

\bibitem[{\citenamefont{Navon et~al.}(2010)\citenamefont{Navon, Nascimb{\`e}ne,
  Chevy, and Salomon}}]{Salomon2010}
\bibinfo{author}{\bibfnamefont{N.}~\bibnamefont{Navon}},
  \bibinfo{author}{\bibfnamefont{S.}~\bibnamefont{Nascimb{\`e}ne}},
  \bibinfo{author}{\bibfnamefont{F.}~\bibnamefont{Chevy}}, \bibnamefont{and}
  \bibinfo{author}{\bibfnamefont{C.}~\bibnamefont{Salomon}},
  \bibinfo{journal}{Science} \textbf{\bibinfo{volume}{328}},
  \bibinfo{pages}{729} (\bibinfo{year}{2010}).

\bibitem[{\citenamefont{Nascimb\`ene et~al.}(2010)\citenamefont{Nascimb\`ene,
  Navon, Jiang, Chevy, and Salomon}}]{Salomon_Nature2010}
\bibinfo{author}{\bibfnamefont{S.}~\bibnamefont{Nascimb\`ene}},
  \bibinfo{author}{\bibfnamefont{N.}~\bibnamefont{Navon}},
  \bibinfo{author}{\bibfnamefont{K.}~\bibnamefont{Jiang}},
  \bibinfo{author}{\bibfnamefont{F.}~\bibnamefont{Chevy}}, \bibnamefont{and}
  \bibinfo{author}{\bibfnamefont{C.}~\bibnamefont{Salomon}},
  \bibinfo{journal}{Nature (London)} \textbf{\bibinfo{volume}{463}},
  \bibinfo{pages}{1057} (\bibinfo{year}{2010}).

\bibitem[{\citenamefont{Van~Houcke et~al.}(2012)\citenamefont{Van~Houcke,
  Werner, Kozik, Prokofev, Svistunov, Ku, Sommer, Cheuk, Schirotzek, and
  Zwierlein}}]{Felix}
\bibinfo{author}{\bibfnamefont{K.}~\bibnamefont{Van~Houcke}},
  \bibinfo{author}{\bibfnamefont{F.}~\bibnamefont{Werner}},
  \bibinfo{author}{\bibfnamefont{E.}~\bibnamefont{Kozik}},
  \bibinfo{author}{\bibfnamefont{N.}~\bibnamefont{Prokofev}},
  \bibinfo{author}{\bibfnamefont{B.}~\bibnamefont{Svistunov}},
  \bibinfo{author}{\bibfnamefont{M.~J.~H.} \bibnamefont{Ku}},
  \bibinfo{author}{\bibfnamefont{A.~T.} \bibnamefont{Sommer}},
  \bibinfo{author}{\bibfnamefont{L.~W.} \bibnamefont{Cheuk}},
  \bibinfo{author}{\bibfnamefont{A.}~\bibnamefont{Schirotzek}},
  \bibnamefont{and} \bibinfo{author}{\bibfnamefont{M.~W.}
  \bibnamefont{Zwierlein}}, \bibinfo{journal}{Nature Phys.}
  \textbf{\bibinfo{volume}{8}}, \bibinfo{pages}{366} (\bibinfo{year}{2012}).

\bibitem[{\citenamefont{Ku et~al.}(2012)\citenamefont{Ku, Sommer, Cheuk, and
  Zwierlein}}]{Zwierlein}
\bibinfo{author}{\bibfnamefont{M.~J.~H.} \bibnamefont{Ku}},
  \bibinfo{author}{\bibfnamefont{A.~T.} \bibnamefont{Sommer}},
  \bibinfo{author}{\bibfnamefont{L.~W.} \bibnamefont{Cheuk}}, \bibnamefont{and}
  \bibinfo{author}{\bibfnamefont{M.~W.} \bibnamefont{Zwierlein}},
  \bibinfo{journal}{Science} \textbf{\bibinfo{volume}{335}},
  \bibinfo{pages}{563} (\bibinfo{year}{2012}).

\bibitem[{\citenamefont{Kohstall et~al.}(2011)\citenamefont{Kohstall, Riedl,
  {S\'{a}nchez Guajardo}, Sidorenkov, {Hecker Denschlag}, and
  Grimm}}]{Kohstall2011}
\bibinfo{author}{\bibfnamefont{C.}~\bibnamefont{Kohstall}},
  \bibinfo{author}{\bibfnamefont{S.}~\bibnamefont{Riedl}},
  \bibinfo{author}{\bibfnamefont{E.~R.} \bibnamefont{{S\'{a}nchez Guajardo}}},
  \bibinfo{author}{\bibfnamefont{L.~A.} \bibnamefont{Sidorenkov}},
  \bibinfo{author}{\bibfnamefont{J.}~\bibnamefont{{Hecker Denschlag}}},
  \bibnamefont{and} \bibinfo{author}{\bibfnamefont{R.}~\bibnamefont{Grimm}},
  \bibinfo{journal}{New Journal of Physics} \textbf{\bibinfo{volume}{13}},
  \bibinfo{pages}{065027} (\bibinfo{year}{2011}).

\bibitem[{\citenamefont{Altman et~al.}(2004)\citenamefont{Altman, Demler, and
  Lukin}}]{Altman2004}
\bibinfo{author}{\bibfnamefont{E.}~\bibnamefont{Altman}},
  \bibinfo{author}{\bibfnamefont{E.}~\bibnamefont{Demler}}, \bibnamefont{and}
  \bibinfo{author}{\bibfnamefont{M.~D.} \bibnamefont{Lukin}},
  \bibinfo{journal}{Phys. Rev. A} \textbf{\bibinfo{volume}{70}},
  \bibinfo{pages}{013603} (\bibinfo{year}{2004}).

\bibitem[{\citenamefont{Carusotto and Castin}(2005)}]{CarusottoCastin2005}
\bibinfo{author}{\bibfnamefont{I.}~\bibnamefont{Carusotto}} \bibnamefont{and}
  \bibinfo{author}{\bibfnamefont{Y.}~\bibnamefont{Castin}},
  \bibinfo{journal}{Phys. Rev. Lett.} \textbf{\bibinfo{volume}{94}},
  \bibinfo{pages}{223202} (\bibinfo{year}{2005}).

\bibitem[{\citenamefont{Bardeen et~al.}(1957)\citenamefont{Bardeen, Cooper, and
  Schrieffer}}]{BCS1957}
\bibinfo{author}{\bibfnamefont{J.}~\bibnamefont{Bardeen}},
  \bibinfo{author}{\bibfnamefont{L.}~\bibnamefont{Cooper}}, \bibnamefont{and}
  \bibinfo{author}{\bibfnamefont{J.}~\bibnamefont{Schrieffer}},
  \bibinfo{journal}{Physical Review} \textbf{\bibinfo{volume}{108}},
  \bibinfo{pages}{5} (\bibinfo{year}{1957}).

\bibitem[{\citenamefont{Anderson}(1958)}]{Anderson1958}
\bibinfo{author}{\bibfnamefont{P.}~\bibnamefont{Anderson}},
  \bibinfo{journal}{Physical Review} \textbf{\bibinfo{volume}{112}},
  \bibinfo{pages}{1900} (\bibinfo{year}{1958}).

\bibitem[{\citenamefont{Blaizot and Ripka}(1985)}]{Ripka1985}
\bibinfo{author}{\bibfnamefont{J.-P.} \bibnamefont{Blaizot}} \bibnamefont{and}
  \bibinfo{author}{\bibfnamefont{G.}~\bibnamefont{Ripka}},
  \emph{\bibinfo{title}{{Quantum Theory of Finite Systems}}}
  (\bibinfo{publisher}{MIT Press}, \bibinfo{address}{Cambridge, Massachusetts},
  \bibinfo{year}{1985}).

\bibitem[{\citenamefont{Castin and Dum}(1998)}]{CastinDum1998}
\bibinfo{author}{\bibfnamefont{Y.}~\bibnamefont{Castin}} \bibnamefont{and}
  \bibinfo{author}{\bibfnamefont{R.}~\bibnamefont{Dum}},
  \bibinfo{journal}{Physical Review A} \textbf{\bibinfo{volume}{57}},
  \bibinfo{pages}{3008} (\bibinfo{year}{1998}).

\bibitem[{\citenamefont{Sinatra et~al.}(2013)\citenamefont{Sinatra, Castin, and
  Witkowska}}]{EPL}
\bibinfo{author}{\bibfnamefont{A.}~\bibnamefont{Sinatra}},
  \bibinfo{author}{\bibfnamefont{Y.}~\bibnamefont{Castin}}, \bibnamefont{and}
  \bibinfo{author}{\bibfnamefont{E.}~\bibnamefont{Witkowska}},
  \bibinfo{journal}{Europhys. Lett.} \textbf{\bibinfo{volume}{102}},
  \bibinfo{pages}{40001} (\bibinfo{year}{2013}).

\bibitem[{\citenamefont{Mora and Castin}(2003)}]{CastinMora2003}
\bibinfo{author}{\bibfnamefont{C.}~\bibnamefont{Mora}} \bibnamefont{and}
  \bibinfo{author}{\bibfnamefont{Y.}~\bibnamefont{Castin}},
  \bibinfo{journal}{Phys. Rev. A} \textbf{\bibinfo{volume}{67}},
  \bibinfo{pages}{053615} (\bibinfo{year}{2003}).

\bibitem[{\citenamefont{Castin}(2004)}]{LesHouches2004}
\bibinfo{author}{\bibfnamefont{Y.}~\bibnamefont{Castin}}, \bibinfo{journal}{J.
  Phys. IV France} \textbf{\bibinfo{volume}{116}}, \bibinfo{pages}{89}
  (\bibinfo{year}{2004}).

\bibitem[{\citenamefont{Burovski et~al.}(2006)\citenamefont{Burovski,
  Prokof'ev, Svistunov, and Troyer}}]{Burovski2006}
\bibinfo{author}{\bibfnamefont{E.}~\bibnamefont{Burovski}},
  \bibinfo{author}{\bibfnamefont{N.}~\bibnamefont{Prokof'ev}},
  \bibinfo{author}{\bibfnamefont{B.}~\bibnamefont{Svistunov}},
  \bibnamefont{and} \bibinfo{author}{\bibfnamefont{M.}~\bibnamefont{Troyer}},
  \bibinfo{journal}{New J. Phys.} \textbf{\bibinfo{volume}{8}},
  \bibinfo{pages}{153} (\bibinfo{year}{2006}).

\bibitem[{\citenamefont{Pricoupenko and Castin}(2007)}]{Pricoupenko2007}
\bibinfo{author}{\bibfnamefont{L.}~\bibnamefont{Pricoupenko}} \bibnamefont{and}
  \bibinfo{author}{\bibfnamefont{Y.}~\bibnamefont{Castin}},
  \bibinfo{journal}{J. Phys. A} \textbf{\bibinfo{volume}{40}},
  \bibinfo{pages}{12863} (\bibinfo{year}{2007}).

\bibitem[{\citenamefont{Juillet}(2007)}]{Juillet2007}
\bibinfo{author}{\bibfnamefont{O.}~\bibnamefont{Juillet}},
  \bibinfo{journal}{New J. Phys.} \textbf{\bibinfo{volume}{9}},
  \bibinfo{pages}{163} (\bibinfo{year}{2007}).

\bibitem[{\citenamefont{Nadal}(2008)}]{Nadal}
\bibinfo{author}{\bibfnamefont{C.}~\bibnamefont{Nadal}},
  \bibinfo{journal}{master internship at ENS (Paris)}  (\bibinfo{year}{2008}).

\bibitem[{\citenamefont{Leggett}(1980)}]{Leggett1980}
\bibinfo{author}{\bibfnamefont{A.}~\bibnamefont{Leggett}},
  \bibinfo{journal}{Journal de physique Colloq.} \textbf{\bibinfo{volume}{41}},
  \bibinfo{pages}{C7\hyphen19} (\bibinfo{year}{1980}).

\bibitem[{\citenamefont{Engelbrecht et~al.}(1997)\citenamefont{Engelbrecht,
  Randeria, and S\'ade~Melo}}]{Engelbrecht1997}
\bibinfo{author}{\bibfnamefont{J.~R.} \bibnamefont{Engelbrecht}},
  \bibinfo{author}{\bibfnamefont{M.}~\bibnamefont{Randeria}}, \bibnamefont{and}
  \bibinfo{author}{\bibfnamefont{C.~A.~R.} \bibnamefont{S\'ade~Melo}},
  \bibinfo{journal}{Phys. Rev. B} \textbf{\bibinfo{volume}{55}},
  \bibinfo{pages}{15153} (\bibinfo{year}{1997}).

\bibitem[{\citenamefont{Randeria}(1995)}]{RanderiaGreenBook}
\bibinfo{author}{\bibfnamefont{M.}~\bibnamefont{Randeria}}, in
  \emph{\bibinfo{booktitle}{Bose-Einstein Condensation}}, edited by
  \bibinfo{editor}{\bibfnamefont{A.}~\bibnamefont{Griffin}},
  \bibinfo{editor}{\bibfnamefont{D.~W.} \bibnamefont{Snoke}}, \bibnamefont{and}
  \bibinfo{editor}{\bibfnamefont{S.}~\bibnamefont{Stringari}}
  (\bibinfo{publisher}{Cambridge University Press}, \bibinfo{address}{Bologna,
  Italia}, \bibinfo{year}{1995}).

\bibitem[{\citenamefont{Marini et~al.}(1998)\citenamefont{Marini, Pistolesi,
  and Strinati}}]{Strinati1998}
\bibinfo{author}{\bibfnamefont{M.}~\bibnamefont{Marini}},
  \bibinfo{author}{\bibfnamefont{F.}~\bibnamefont{Pistolesi}},
  \bibnamefont{and} \bibinfo{author}{\bibfnamefont{G.}~\bibnamefont{Strinati}},
  \bibinfo{journal}{European Physical Journal B} \textbf{\bibinfo{volume}{1}},
  \bibinfo{pages}{151} (\bibinfo{year}{1998}).

\bibitem[{\citenamefont{Belzig et~al.}(2007)\citenamefont{Belzig, Schroll, and
  Bruder}}]{Belzig2007}
\bibinfo{author}{\bibfnamefont{W.}~\bibnamefont{Belzig}},
  \bibinfo{author}{\bibfnamefont{C.}~\bibnamefont{Schroll}}, \bibnamefont{and}
  \bibinfo{author}{\bibfnamefont{C.}~\bibnamefont{Bruder}},
  \bibinfo{journal}{Phys. Rev. A} \textbf{\bibinfo{volume}{75}},
  \bibinfo{pages}{063611} (\bibinfo{year}{2007}).

\bibitem[{\citenamefont{Pieri and Strinati}(2000)}]{Pieri2000}
\bibinfo{author}{\bibfnamefont{P.}~\bibnamefont{Pieri}} \bibnamefont{and}
  \bibinfo{author}{\bibfnamefont{G.}~\bibnamefont{Strinati}},
  \bibinfo{journal}{Physical Review B} \textbf{\bibinfo{volume}{61}},
  \bibinfo{pages}{15370} (\bibinfo{year}{2000}).

\bibitem[{\citenamefont{Petrov et~al.}(2004)\citenamefont{Petrov, Salomon, and
  Shlyapnikov}}]{Petrov2004}
\bibinfo{author}{\bibfnamefont{D.~S.} \bibnamefont{Petrov}},
  \bibinfo{author}{\bibfnamefont{C.}~\bibnamefont{Salomon}}, \bibnamefont{and}
  \bibinfo{author}{\bibfnamefont{G.~V.} \bibnamefont{Shlyapnikov}},
  \bibinfo{journal}{Physical Review Letters} \textbf{\bibinfo{volume}{93}},
  \bibinfo{pages}{090404} (\bibinfo{year}{2004}).

\bibitem[{\citenamefont{Brodsky et~al.}(2006)\citenamefont{Brodsky, Kagan,
  Klaptsov, Combescot, and Leyronas}}]{Leyronas2006}
\bibinfo{author}{\bibfnamefont{I.~V.} \bibnamefont{Brodsky}},
  \bibinfo{author}{\bibfnamefont{M.~Y.} \bibnamefont{Kagan}},
  \bibinfo{author}{\bibfnamefont{A.~V.} \bibnamefont{Klaptsov}},
  \bibinfo{author}{\bibfnamefont{R.}~\bibnamefont{Combescot}},
  \bibnamefont{and} \bibinfo{author}{\bibfnamefont{X.}~\bibnamefont{Leyronas}},
  \bibinfo{journal}{Physical Review A} \textbf{\bibinfo{volume}{73}},
  \bibinfo{pages}{032724} (\bibinfo{year}{2006}).

\bibitem[{\citenamefont{Astrakharchik et~al.}(2007)\citenamefont{Astrakharchik,
  Combescot, and Pitaevskii}}]{Combescot2007}
\bibinfo{author}{\bibfnamefont{G.~E.} \bibnamefont{Astrakharchik}},
  \bibinfo{author}{\bibfnamefont{R.}~\bibnamefont{Combescot}},
  \bibnamefont{and} \bibinfo{author}{\bibfnamefont{L.~P.}
  \bibnamefont{Pitaevskii}}, \bibinfo{journal}{Phys. Rev. A}
  \textbf{\bibinfo{volume}{76}}, \bibinfo{pages}{063616}
  (\bibinfo{year}{2007}).

\bibitem[{\citenamefont{Leggett}(2006)}]{Leggett2006}
\bibinfo{author}{\bibfnamefont{A.~J.} \bibnamefont{Leggett}},
  \emph{\bibinfo{title}{{Quantum Liquids}}} (\bibinfo{publisher}{Oxford
  University Press}, \bibinfo{address}{Oxford}, \bibinfo{year}{2006}).

\bibitem[{\citenamefont{Carruther and Nieto}(1968)}]{Nieto1968}
\bibinfo{author}{\bibfnamefont{P.}~\bibnamefont{Carruther}} \bibnamefont{and}
  \bibinfo{author}{\bibfnamefont{M.~M.} \bibnamefont{Nieto}},
  \bibinfo{journal}{Rev. Mod. Phys.} \textbf{\bibinfo{volume}{40}},
  \bibinfo{pages}{411} (\bibinfo{year}{1968}).

\bibitem[{\citenamefont{Castin and Sinatra}(2013)}]{CastinSinatra2012}
\bibinfo{author}{\bibfnamefont{Y.}~\bibnamefont{Castin}} \bibnamefont{and}
  \bibinfo{author}{\bibfnamefont{A.}~\bibnamefont{Sinatra}}, in
  \emph{\bibinfo{booktitle}{{Physics of Quantum Fluids: New Trends and Hot
  Topics in Atomic and Polariton Condensates}}}, edited by
  \bibinfo{editor}{\bibfnamefont{M.}~\bibnamefont{Modugno}} \bibnamefont{and}
  \bibinfo{editor}{\bibfnamefont{A.}~\bibnamefont{Bramati}}
  (\bibinfo{publisher}{Springer-Verlag, Berlin, Germany},
  \bibinfo{year}{2013}), vol. \bibinfo{volume}{177 Springer Series in
  Solid-State Sciences}.

\bibitem[{\citenamefont{Gritsev et~al.}(2008)\citenamefont{Gritsev, Demler, and
  Polkovnikov}}]{Gritsev2008}
\bibinfo{author}{\bibfnamefont{V.}~\bibnamefont{Gritsev}},
  \bibinfo{author}{\bibfnamefont{E.}~\bibnamefont{Demler}}, \bibnamefont{and}
  \bibinfo{author}{\bibfnamefont{A.}~\bibnamefont{Polkovnikov}},
  \bibinfo{journal}{Phys. Rev. A} \textbf{\bibinfo{volume}{78}},
  \bibinfo{pages}{063624} (\bibinfo{year}{2008}).

\bibitem[{\citenamefont{Kitagawa et~al.}(2011)\citenamefont{Kitagawa, Aspect,
  Greiner, and Demler}}]{AspectDemler2011}
\bibinfo{author}{\bibfnamefont{T.}~\bibnamefont{Kitagawa}},
  \bibinfo{author}{\bibfnamefont{A.}~\bibnamefont{Aspect}},
  \bibinfo{author}{\bibfnamefont{M.}~\bibnamefont{Greiner}}, \bibnamefont{and}
  \bibinfo{author}{\bibfnamefont{E.}~\bibnamefont{Demler}},
  \bibinfo{journal}{Phys. Rev. Lett.} \textbf{\bibinfo{volume}{106}},
  \bibinfo{pages}{115302} (\bibinfo{year}{2011}).

\bibitem[{\citenamefont{Greiner et~al.}(2005)\citenamefont{Greiner, Regal,
  Stewart, and Jin}}]{GreinerRegal2005}
\bibinfo{author}{\bibfnamefont{M.}~\bibnamefont{Greiner}},
  \bibinfo{author}{\bibfnamefont{C.~A.} \bibnamefont{Regal}},
  \bibinfo{author}{\bibfnamefont{J.~T.} \bibnamefont{Stewart}},
  \bibnamefont{and} \bibinfo{author}{\bibfnamefont{D.~S.} \bibnamefont{Jin}},
  \bibinfo{journal}{Phys. Rev. Lett.} \textbf{\bibinfo{volume}{94}},
  \bibinfo{pages}{110401} (\bibinfo{year}{2005}).

\bibitem[{\citenamefont{Rom et~al.}(2006)\citenamefont{Rom, Best, van Oosten,
  Schneider, F\"{o}lling, Paredes, and Bloch}}]{Bloch2006}
\bibinfo{author}{\bibfnamefont{T.}~\bibnamefont{Rom}},
  \bibinfo{author}{\bibfnamefont{T.}~\bibnamefont{Best}},
  \bibinfo{author}{\bibfnamefont{D.}~\bibnamefont{van Oosten}},
  \bibinfo{author}{\bibfnamefont{U.}~\bibnamefont{Schneider}},
  \bibinfo{author}{\bibfnamefont{S.}~\bibnamefont{F\"{o}lling}},
  \bibinfo{author}{\bibfnamefont{B.}~\bibnamefont{Paredes}}, \bibnamefont{and}
  \bibinfo{author}{\bibfnamefont{I.}~\bibnamefont{Bloch}},
  \bibinfo{journal}{Nature} \textbf{\bibinfo{volume}{444}},
  \bibinfo{pages}{733} (\bibinfo{year}{2006}).

\end{thebibliography}


\end{document}